\documentclass{aa}  
\usepackage{graphicx}
\usepackage{txfonts}
\usepackage{xcolor}
\usepackage{multirow}
\usepackage{hyperref}

\begin{document}

   \title{Unveiling the origin of \textit{XMM-Newton} soft proton flares}

   \subtitle{I. Design and validation of a response matrix for proton spectral analysis}

   \author{V. Fioretti
          \inst{1}\fnmsep\thanks{\email{valentina.fioretti@inaf.it}},
          T. Mineo
          \inst{2},
          S. Lotti
          \inst{3},
          S. Molendi
         \inst{4},
          G. Lanzuisi
          \inst{1},
          R. Amato
          \inst{5},
          C. Macculi
          \inst{3},
        M. Cappi
        \inst{1},
         M. Dadina
        \inst{1},
         S. Ettori
        \inst{1},
         F. Gastaldello
        \inst{4}
          }

   \institute{INAF, Osservatorio di Astrofisica e Scienza dello Spazio (OAS) di Bologna, via P. Gobetti 93/3, 40129 Bologna, Italy
         \and
             INAF, Istituto di Astrofisica Spaziale e Fisica Cosmica (IASF) di Palermo, Via Ugo La Malfa 153, 90146 Palermo, Italy
        \and
             INAF, Istituto di Astrofisica e Planetologia Spaziali (IAPS), Via del Fosso del Cavaliere 100, 00133 Roma, Italy
       \and
       INAF, Istituto di Astrofisica Spaziale e Fisica Cosmica (IASF) di Milano, Via E. Bassini 15, 20133 Milano, Italy
       \and
       INAF, Osservatorio Astronomico di Roma (OAR), Via Frascati 33, 00078 Monte Porzio Catone (RM), Italy
             }

   \date{Received September 15, 1996; accepted March 16, 1997}

 
  \abstract
   {Low-energy ($<$ 300 keV) protons entering the field of view of \textit{XMM-Newton} can scatter with the X-ray mirror surface and reach the focal plane. They are observed in the form of a sudden increase in the background level, the so-called soft proton flares, affecting up to 40\% of the mission observing time. Soft protons can hardly be disentangled from true X-ray events and cannot be rejected on board.}
   {All future high throughput grazing incidence X-ray telescopes operating outside the radiation belts are potentially affected by soft proton-induced contamination that must be foreseen and limited since the design phase. In-flight \textit{XMM-Newton}'s observations of soft protons represent a unique laboratory to validate and improve our understanding of their interaction with the mirror, optical filters, and X-ray instruments. At the same time, such models would link the observed background flares to the primary proton population encountered by the telescope, converting \textit{XMM-Newton} into a monitor for soft protons.
   
   }
   {We built a Geant4 simulation of \textit{XMM-Newton}, including a verified mass model of the X-ray mirror, the focal plane assembly, and the EPIC MOS and pn-CCDs. Analytical computations and, when available, laboratory measurements collected from literature were used to verify the correct modelling of the proton scattering and transmission to the detection plane.
   Similarly to the instrument X-ray response, we encoded the energy redistribution and proton transmission efficiency into a redistribution matrix file (RMF), mapping the probability that a proton from 2 to 300 keV is detected in a certain detector channel, and an auxiliary response file (ARF), storing the grasp towards protons. Both files were formatted according to the standard NASA calibration database and any compliant X-ray data analysis tool can be used to simulate or analyse soft proton-induced background spectra. An overall systematic uncertainty of 30\% was assumed on the basis of the estimated accuracy of the mirror geometry and transmission models.
   
   }
   {
   For the validation, three averaged soft proton spectra, one for each filter configuration, were extracted from a collection of 13 years of MOS observations of the focused non X-ray background and analysed with \textit{Xspec}.
   A similar power-law distribution is found for the three filter configurations, plus black-body-like emission below tens of keV used as a correction factor, based on the dedicated spectral analysis of 55 in-flight proton flares presented in Paper II.
   The best-fit model is in agreement with the power-law distribution predicted from independent measurements for the \textit{XMM-Newton} orbit, spent mostly in the magnetosheath and nearby regions. For the first time we are able to link detected soft proton flares with the proton radiation environment in the Earth's magnetosphere, while proving the validity of the simulation chain in predicting the background of future missions. 
   Benefiting from this work and contributions from the \textit{Athena} instrument consortia, we also present the response files for the \textit{Athena} mission and updated estimates for its focused charged background.  
   
   
   }
   {}


\titlerunning{Unveiling the origin of \textit{XMM-Newton} soft proton flares I}
\authorrunning{V. Fioretti et al.}

   \maketitle
%

\section{Introduction}
X-ray space telescopes based on focusing mirrors exploit the grazing angle reflection in the $<80$ keV band to concentrate photons into a smaller detection surface, reducing the contamination by non X-ray background (NXB). 
Most X-ray missions (\textit{XMM-Newton}, \textit{Chandra}, eROSITA, Swift/XRT, NuSTAR, and XRISM) use a set of nested Wolter-I \citep{2021hai4.book....3P} shells, requiring two reflections in paraboloid and hyperboloid confocal and coaxial mirrors. The ESA \textit{XMM-Newton} X-ray telescope, launched in 1999, still features the largest full band effective area \citep{refId0, xrism_quickref}, with 1500 cm$^{2}$ at 1 keV  and 900 cm$^{2}$ at 7 keV \citep{2002SPIE.4496....8A}. Its electroforming nickel X-ray mirrors, lighter than the NASA \textit{Chandra} mission's polished glass shells but with a larger figure error, allowed 58 shells to be packed in each mirror module against \textit{Chandra}'s four \citep{10.1117/12.278890}. 
In its highly eccentric orbit, reaching at the time of writing\footnote{`\textit{XMM-Newton} Users Handbook', Issue 2.21, 2023} an apogee $>9\times10^{4}$ km, \textit{XMM-Newton} spends $>80\%$ of the operational time outside the radiation belts and the shielding provided by the Earth's magnetic field. This, combined with the high throughput of the optics, is the cause of the high contamination by the so-called soft proton flares during astronomical observations. 
\\
Low-energy protons ($< 300$ keV) populate both the interplanetary space and the outer magnetosphere, including the Earth’s magnetotail \citep{Lotti2018} in the form of a dynamic and persistent flux characterised by both short- and long-term variability induced by transient shock waves (for example coronal mass ejections) and the solar activity cycle. They can enter the field of view of X-ray-focusing optics, scatter with the mirror surface, and reach the focal plane, as firstly discovered by the degradation of the front-illuminated \textit{Chandra} ACIS charge-coupled devices (CCDs) during the first radiation belt passages \citep{2000SPIE.4140...99O}. Onboard protecting measures are in place for both \textit{Chandra} and \textit{XMM-Newton} to avoid further radiation damage from trapped particles. However, soft protons encountered along the orbit can still reach the focal plane and produce energy deposits that can hardly be disentangled from true X-ray events. These soft proton flares are sudden and increase unpredictably the background rate. In \textit{XMM-Newton} they can prevail over the quiescent background level up to 1000\% \citep{2007A&A...464.1155C}, with a duration lasting from $\sim100$ seconds to hours \citep{del04} and the loss of up to 40\% of observing time \citep{2017ExA....44..297M}.
In contrast, they only affect 8\% of the \textit{Chandra} observation time while eROSITA, operating in the second Lagrange point in the direction opposite to the Sun, only rarely observed significant flares and with a shorter (tens of seconds) duration \citep{2020SPIE11444E..1OF}. 
\\
All future high throughput grazing incidence X-ray telescopes operating outside the radiation belts (for example the future ESA X-ray large mission \textit{Athena}\footnote{https://www.cosmos.esa.int/web/athena}) will be potentially affected by soft proton-induced contamination that must be foreseen and limited since the design phase. 
The simulation of the soft proton-induced background requires a dedicated modelling of the radiation environment encountered by the spacecraft, the proton transmission efficiency through the mirror module, filters and its final interaction with the detection system. The validation of the simulation chain can only be obtained by comparison of the predicted background flux with in-flight data. In this context, \textit{XMM-Newton} provided an unprecedented archive of soft proton flare spectra and represents a unique laboratory for the characterisation of the detection response towards low-energy protons entering the field of view. 
\\
We present the design and validation of a proton response matrix for the \textit{XMM-Newton} telescope (Sect. \ref{sec:build}) and the \textit{Athena} proposal (before descoping, Sect. \ref{sec:athena}). The soft proton flares are the result of the convolution of the incoming proton flux at the telescope pupil with the response matrix. Our goals validate the simulation chain against in-flight data, speeding up the background prediction for future missions, and the X-ray analysis of \textit{XMM-Newton}'s soft proton flares to derive the spectral models for the protons encountered along the orbit, improving our knowledge of the near-Earth radiation environment.
\section{Design of a response matrix for protons}\label{sec:resp}
A preliminary simulation of the soft proton-induced background, using a simplified model of the X-ray mirror and the EPIC pn detector (see Sect. \ref{sec:epic}) was published in \citet{2016SPIE.9905E..6WF} but the validation against real data was not possible because of (i) the lack of in-flight data averaged along the spacecraft orbit, (ii) the lack of a precise estimate of the proton environment and (iii) the use of approximated scattering physics models. In recent years, the EXTraS (Exploring the X-ray Transient and variable Sky) project \citep{2017ExA....44..309S} funded by the FP7 European programme, provided an unbiased database of \textit{XMM-Newton} EPIC blank sky observations for the characterisation of the focused charged particle background collected in 13 years of data. 
In parallel, measurements of proton scattering at grazing angles ($<1^{\circ}-2^{\circ}$) on the surface of samples of the eROSITA X-ray mirror \citep{2017SPIE10397E..0WD} and the comparison with simulations \citep{2017ExA....44..413F, 2020ExA....49..115A} have defined with sufficient accuracy the driving physics models for the scattering of protons at grazing incident angles. 
The need for accurate estimates of the particle background on board the \textit{Athena} mission has fostered an in depth study of the soft proton environment in L1 and L2, with the definition of fluxes and spectral distribution for different solar conditions \citep{Lotti2018}. With this improved knowledge of the environment, the physics of the proton interaction, and the in-flight observations, we can now build a verified and optimised simulation of the focused charged particle background.
\\
The mirror simulation is performed with two independent simulation frameworks using the Geant4 toolkit \citep{g4_1, 2016NIMPA.835..186A} and a ray-tracing in order to verify the geometry and physics models and estimate potential systematic effects in their implementation (see Sect. \ref{sec:scatt}). The energy and angular distribution of the protons at a given distance from the focal plane is extracted and given as input to the Geant4 simulation of the focal plane assembly (FPA), including baffles, optical filters, and the detectors (Sect. \ref{sec:epic}). From the FPA simulation, we reconstruct the counts on the EPIC CCDs, applying a pattern flag, according to the instrument read-out configuration. The product of this processing stage is a data level 1 FITS file listing the count energy, position and pattern flags, one for each Geant4 simulation run of each input energy. In the data level 2 we compute for each input energy the energy probability distribution in the instrument channels, normalised to 1, and the grasp. For X-ray photons, that are mono-directional, the response ARF file is the product of the effective area of the mirror, that is, geometric area multiplied by the reflection efficiency, with the filter transmission efficiency and the detector quantum efficiency. For the soft proton environment, we assume an isotropic distribution at the mirror entrance. The simulated grasp is the product of the system efficiency multiplied by the proton aperture solid angle at the mirror entrance, in units of cm$^{2}$ sr. The response matrix is formatted according to the NASA OGIP (Office of Guest Investigators Program) calibration database (caldb) format, and it consists of an RMF and ARF file in FITS (Flexible Image Transport System) format. Any X-ray data analysis tool available to the X-ray astronomy community (for example XSPEC \citep{1996ASPC..101...17A}, SPEX \citep{2024zndo..10822753K}) can be used to simulate the soft proton-induced background spectra, for any given condition of the orbit proton environment without the need to run again the simulation pipeline. Each combination of optical filter and EPIC X-ray instrument is associated to different response files.
Their technical verification and the validation against archival MOS in-flight data is reported in Sect. \ref{sec:vv}.
\\
The X-ray spectral analysis of a sample of MOS and pn soft proton flares, for different seasonal and solar conditions, is presented in \citet{Mineo2024}, hereafter Paper II, to test the general validity of the response files with actual observations.

\section{\textit{XMM-Newton} simulation}\label{sec:xmm_sim}
      \begin{figure*}
   \resizebox{0.99\hsize}{!}
            {\includegraphics[width=0.5\linewidth]{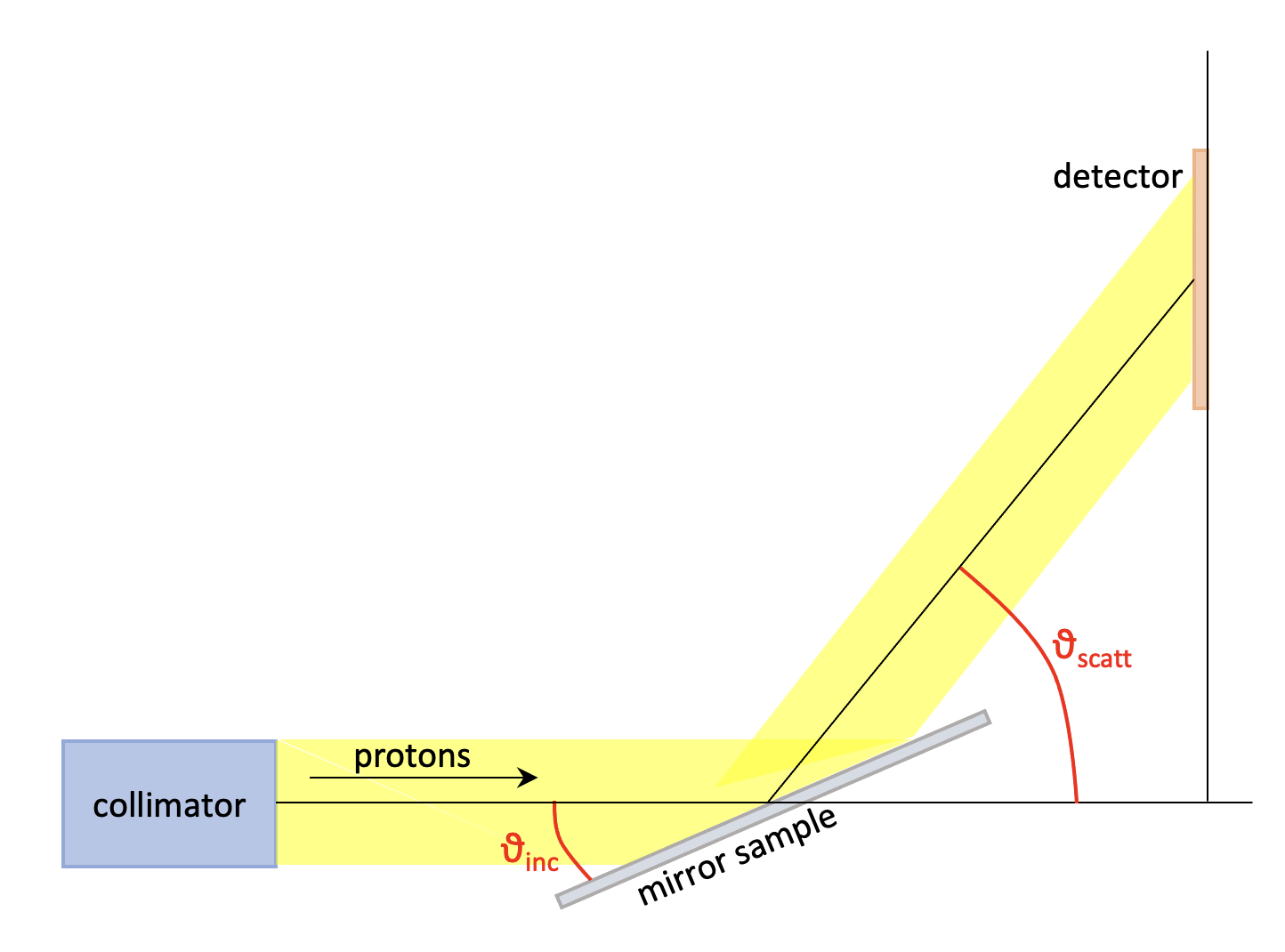}
            \includegraphics[width=0.5\linewidth]{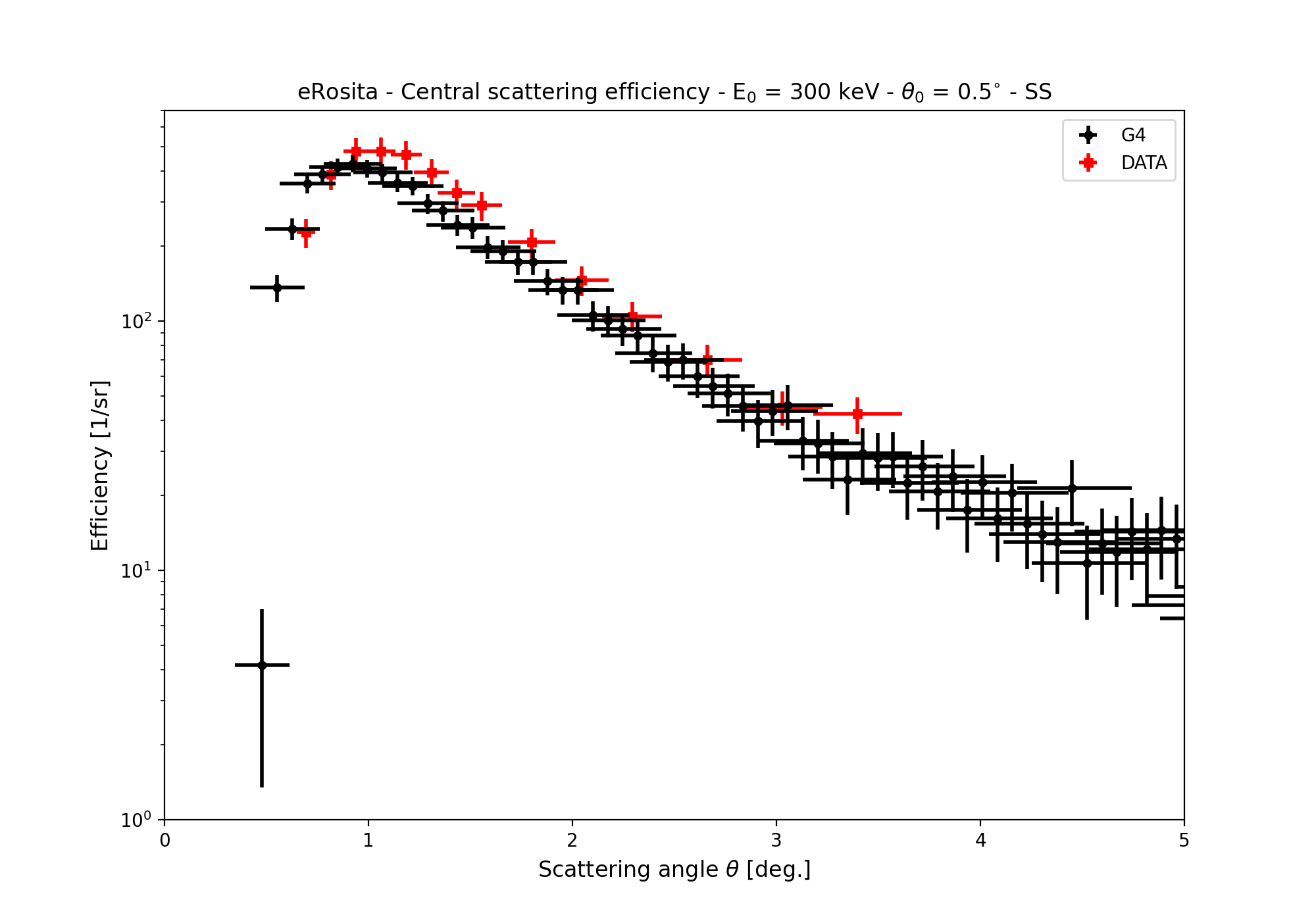}}
            \\
    \resizebox{0.99\hsize}{!}
            {\includegraphics[width=0.5\linewidth]{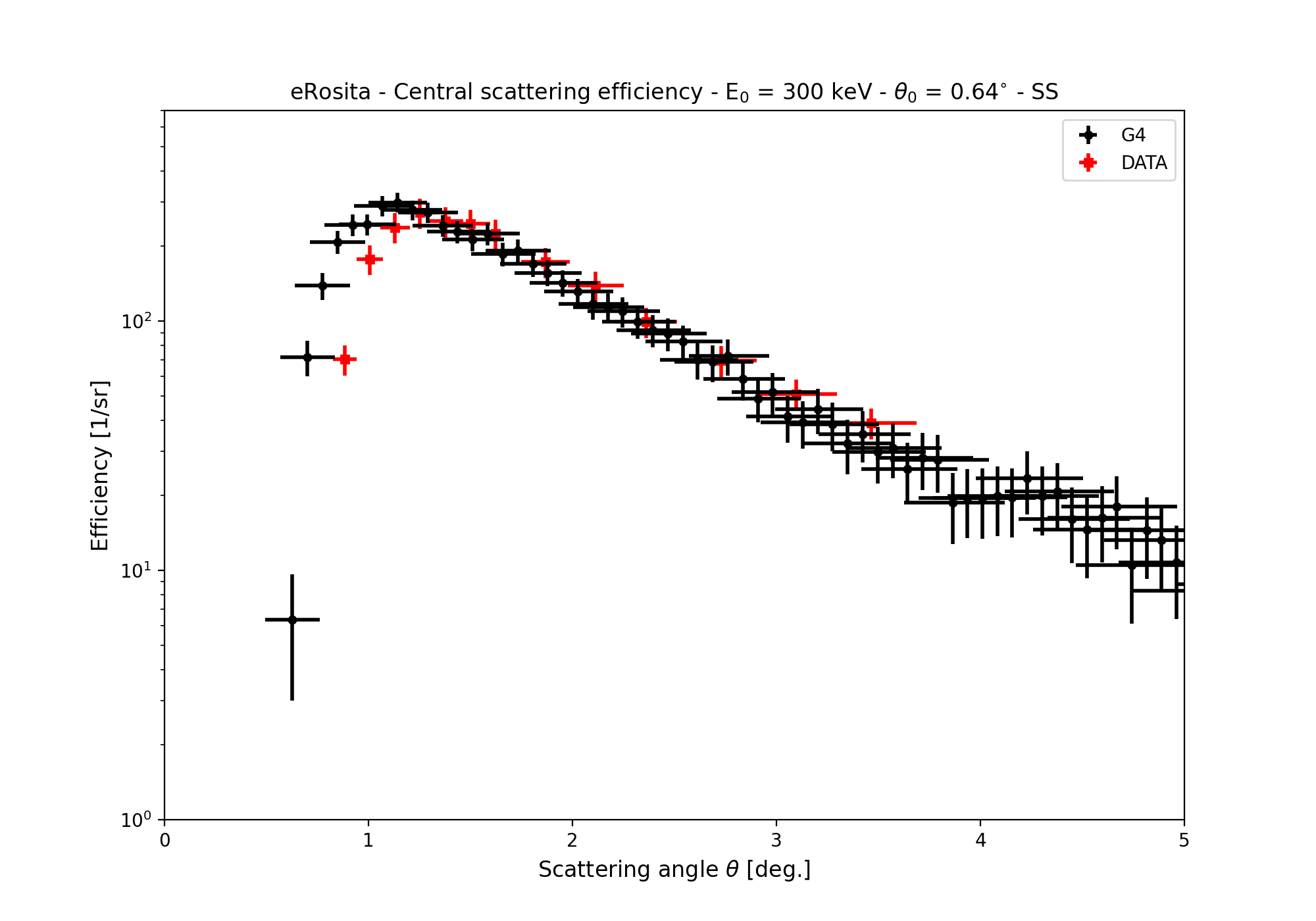}    
            \includegraphics[width=0.5\linewidth]{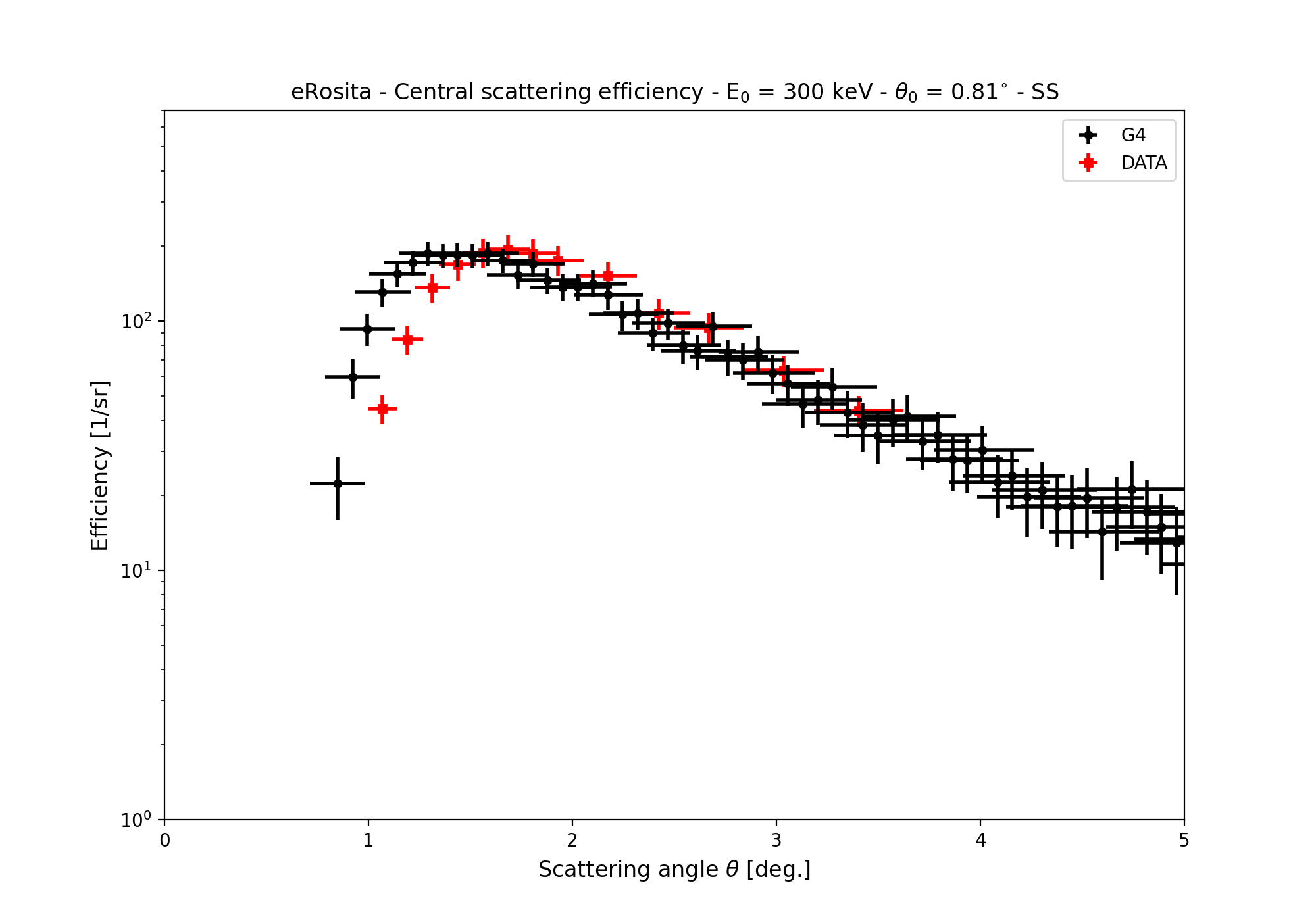}    
            }
      \caption{Proton scattering at grazing angles. (top left) Schematic view of the simulation setup, showing the proton beam exiting the collimator and scattered by the mirror sample towards the detector. Measured, in red, and simulated, in black, proton scattering efficiency at 300 keV for three grazing incident angles, $0.5^{\circ}$ (top right), $0.64^{\circ}$ (bottom left), and $0.81^{\circ}$ (bottom right), hitting the eROSITA mirror sample. The simulation error bars are the sum of the $1\sigma$ statistical and systematic uncertainties.}
         \label{fig:erosita}
   \end{figure*}
\textit{XMM-Newton} carries three Wolter type-I mirror modules composed by 58 Gold-coated Nichel shells \citep{Jansen2001}, with a diameter ranging from $\sim30$ to 70 cm and a focal length of 7.5 m. An X-ray baffle is placed at the entrance of the mirrors for stray-light suppression \citep{1999SPIE.3737..396D}. Its inclusion is mandatory in the mirror simulation because it also reduces the incoming proton flux. The baffle is made of Invar, a Nickel-Iron alloy commonly used for space applications because of its low coefficient of thermal expansion, and it consists of two planes of 59 circular strips and 16 radial spokes at about 10 cm from the mirror entrance.
The spacecraft carries three X-ray European Photon Imaging Cameras (EPIC), each at the focal plane of the mirror modules and sensitive up to $\sim15$ keV. Two of the cameras are MOS (Metal Oxide Semi-conductor) CCD arrays \citep{2001A&A...365L..27T} where the presence of Reflection Grating Spectrometers (RGS) behind the X-ray telescopes shadows about 50\% of the incoming proton flux. The third EPIC camera uses pn CCDs \citep{Struder2001} and no RGS is present. The detectors share the same design for the filter wheel, optical filters, and radiation shielding baffles.
\\
The Geant4 toolkit is an open-source Monte Carlo simulation library for particle and radiation transport through matter. The Geant4-based BoGEMMS framework \citep{2012SPIE.8453E..35B} was used throughout the work to build the mass models and describe the interaction of the protons from the mirror entrance to their absorption in the X-ray detectors. 
   \begin{figure}
   \resizebox{0.9\hsize}{!}
        {\includegraphics[width=0.3\linewidth]{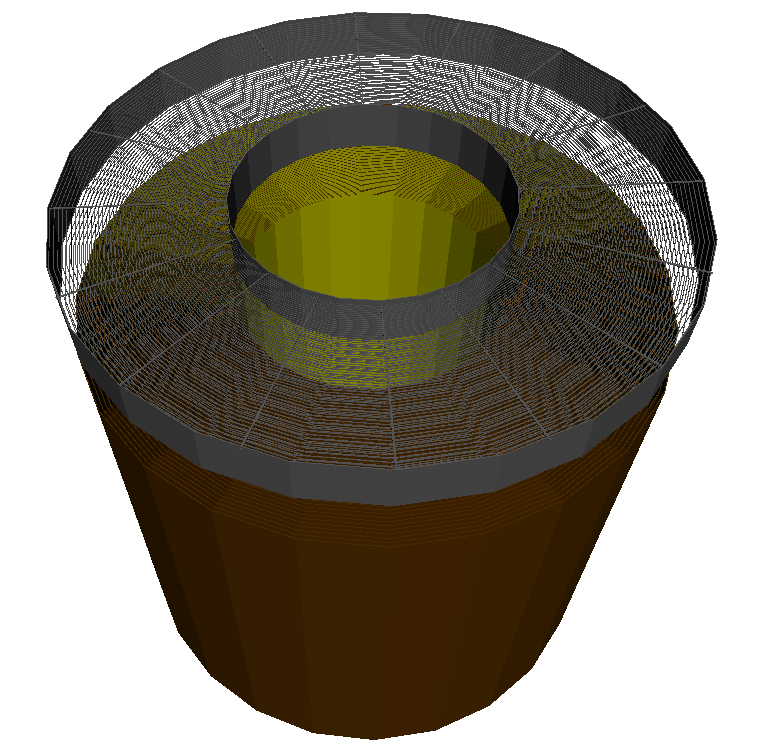}}
            \\
    \resizebox{0.9\hsize}{!}
           {\includegraphics[width=0.5\linewidth]{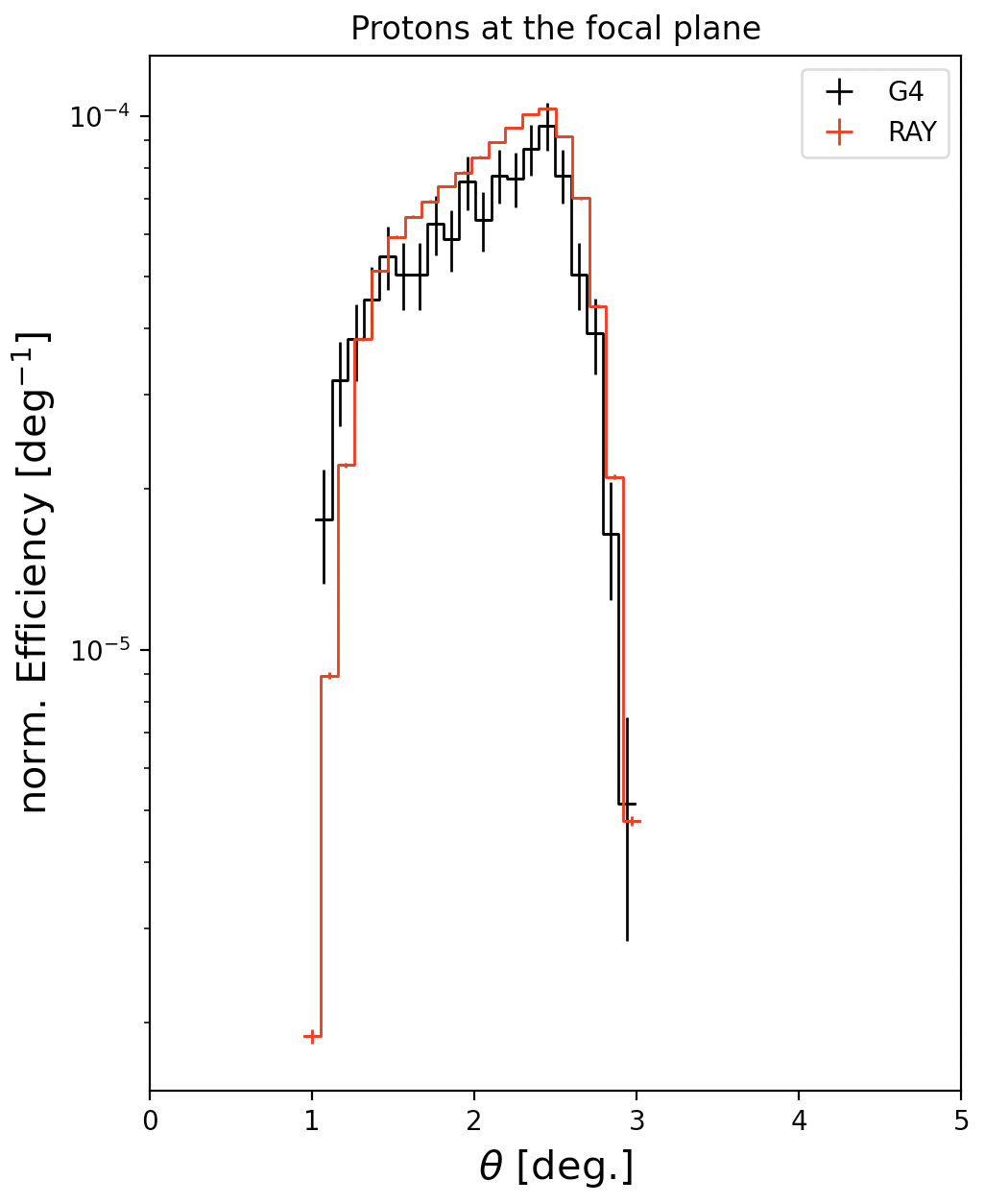}}
      \caption{\textit{XMM-Newton} mirror simulation. (top) Geant4 mass model of the mirror module including the X-ray baffle, in grey, placed at the telescope pupil. (bottom) Distribution of the proton angular distribution at the focal plane, simulated in Geant4 (black) and ray-tracing (red).}
         \label{fig:xmm_mirror}
   \end{figure}

\subsection{Proton scattering}\label{sec:scatt}
The accuracy in reproducing the proton scattering efficiency and angular distribution is fundamental for the simulation of the soft proton-induced background. Because of the required low-energy of the proton beam and the small incident angles ($<1^{\circ}-2^{\circ}$), the first detailed set of scattering measurements was only published in \citet{2015ExA....39..343D} using a sample of the eROSITA mirror shell, composed by a Gold-coated Nickel substrate similar to the \textit{XMM-Newton} design.
A preliminary comparison against the Geant4 simulation \citep{2017ExA....44..413F} showed that the grazing angle reflection of protons is likely due to Coulomb interactions with the electron field of the nuclei. However, the large systematic uncertainties did not allow us to pinpoint the best model between single scattering and the elastic Remizovitch solution.
\\
In the Geant4's single scattering model, each Coulomb collision is simulated, ensuring the needed precision when modelling grazing angle proton scattering, where the proton interacts with the nuclei at the surface and escape after one or more interactions. The only available analytical model describing particle reflected by solids at glancing angles was proposed by \citet{1980JETP...52..225R} but, as studied in \citet{2020ExA....49..115A}, the model depends on the scattering property of the material that can only extracted by fitting the Remizovich formula to a set of measurements. The only exact solution available for the Remizovich equation is in its elastic approximation, with reflection efficiency of 100\% and no energy degradation.
\\
A new data-set was published in \citet{2017SPIE10397E..0WD}, measuring the proton scattering efficiency for 300 keV protons hitting the eROSITA sample at an incident angle of $0.5^{\circ}$, $0.64^{\circ}$ and $0.81^{\circ}$. 
We updated the proton scattering simulations using the new experimental configuration (see \citet{2017ExA....44..413F} for a detailed description of the simulated setup). 
The proton beam hits the eROSITA mirror and a detector is placed at different height to measure the scattering efficiency at different scattering angles, defined as the angle between the plane and the detector position (Fig. \ref{fig:erosita}, top left panel). The number of detected protons, for each position or scattering angle, is normalised for the protons hitting the target and the solid angle subtended by the detector to compute the scattering efficiency, in $sr^{-1}$.
The comparison with the measured scattering efficiency is shown in Fig. \ref{fig:erosita}. All curves peak at specular angles, that is, when $\theta_{scatt} = 2\times\theta_{inc}$, as measured in previous works.
The efficiency predicted by Geant4, using the single scattering model, reproduces with sufficient accuracy the measured efficiency within $1\sigma-2\sigma$ errors, that include the systematic uncertainties of the experimental apparatus, proving the Coulomb interaction as the driving mechanism. The \textit{G4EmStandardPhysicsSS} electromagnetic physics list was used throughout the work. As for the reason of the discrepancies at small scattering angles, we plan to investigate in further studies if modifications from ideal surfaces (for instance micro-roughness, oxidation) can reduce the measured scattering efficiency. 
\\
The Geant4 mass model of the \textit{XMM-Newton} mirror module is shown in Fig. \ref{fig:xmm_mirror} (top panel), with the X-ray baffle in grey, the Nickel shells in red, and the Gold coating in yellow, visible in the inner side. The Wolter-I design is approximated using truncated cones for the paraboloid and hyperboloid sections. We estimated the systematic uncertainty introduced by such approximations comparing the proton transmission through the mirror with independent ray-tracing simulations.
The ray-tracing code is a stand-alone software that uses a Monte Carlo method to assign the particle angular and energy distribution according to a defined analytical model \citep{2017ExA....44..287M}. While this method can implement the exact mirror design and reduces significantly the computational time, it requires an analytical model for the physics interaction. 
The X-ray photon reflectivity, modelled according to tabulated values \citep{HENKE1993181, 1994SPIE.2279..325O} and including the specular direction Gaussian spread induced by the surface micro-roughness, has been verified against the mission official X-ray effective area and vignetting at 1.5 keV and 6.4 keV \citep{2002SPIE.4496....8A}. Discrepancies between the two sets of curves are always lower than 5\%. The verified X-ray optics geometry was then used to simulate in ray-tracing the proton transmission, using the elastic Remizovitch model. The Geant4 implementation of the same elastic model, firstly coded in \citet{2017ExA....44..413F}, was used to verify the X-ray optics mass model against the ray-tracing implementation. We obtained a discrepancy of $10\%$ in the proton transmission caused by the Wolter-I conical approximation. The angular distribution, plotted as a function of the angle from the telescope axis (Fig. \ref{fig:xmm_mirror}, bottom panel) is in good agreement.

\subsection{EPIC simulation}\label{sec:epic}

\begin{figure*}
   \resizebox{\hsize}{!}
            {\includegraphics[width=0.27\linewidth]{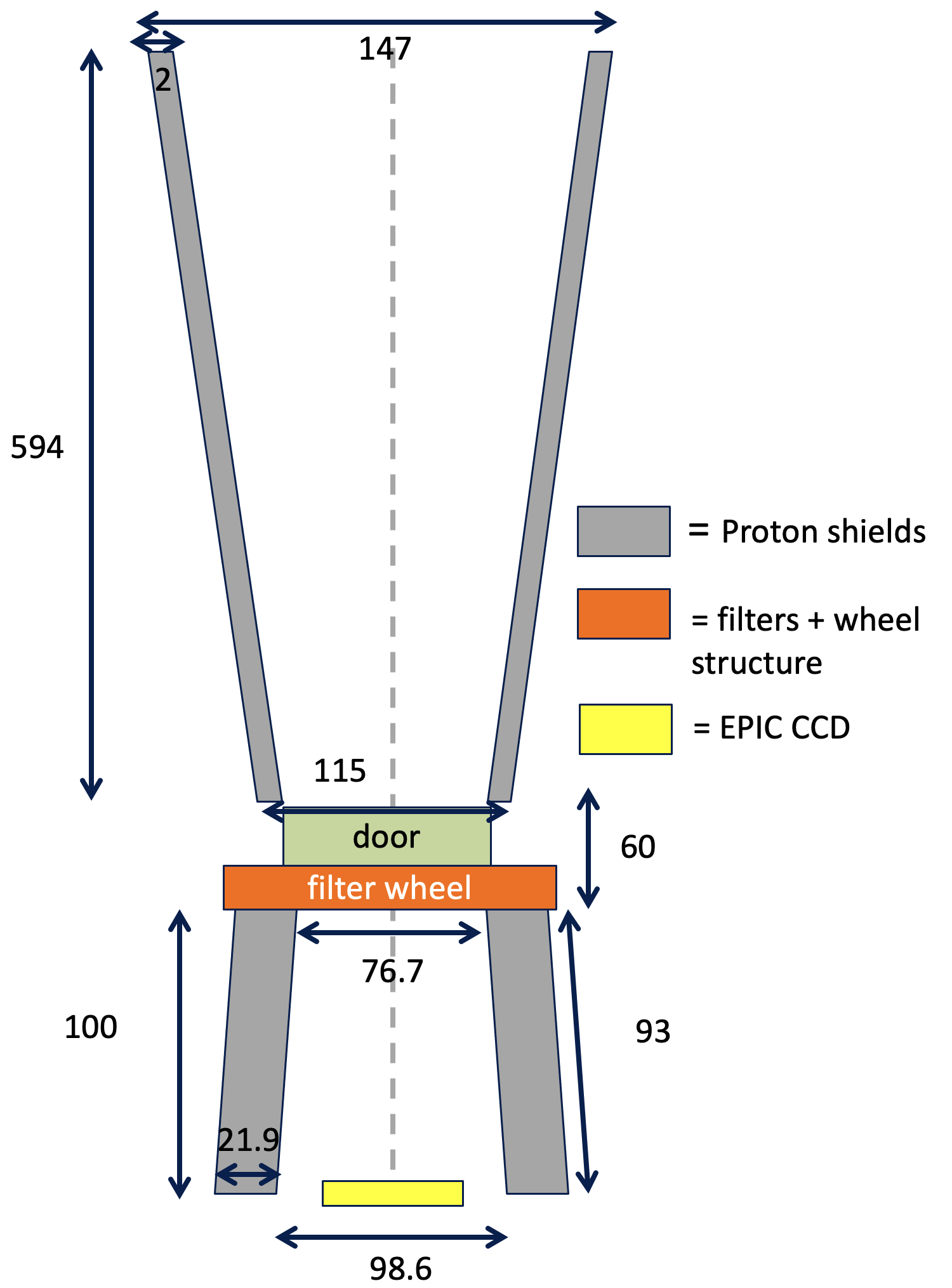}
            \includegraphics[width=0.23\linewidth]{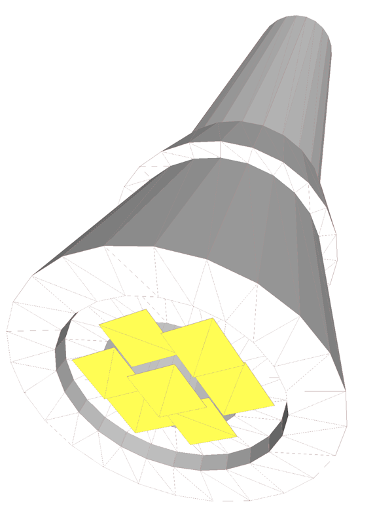}
            \includegraphics[width=0.4\linewidth]{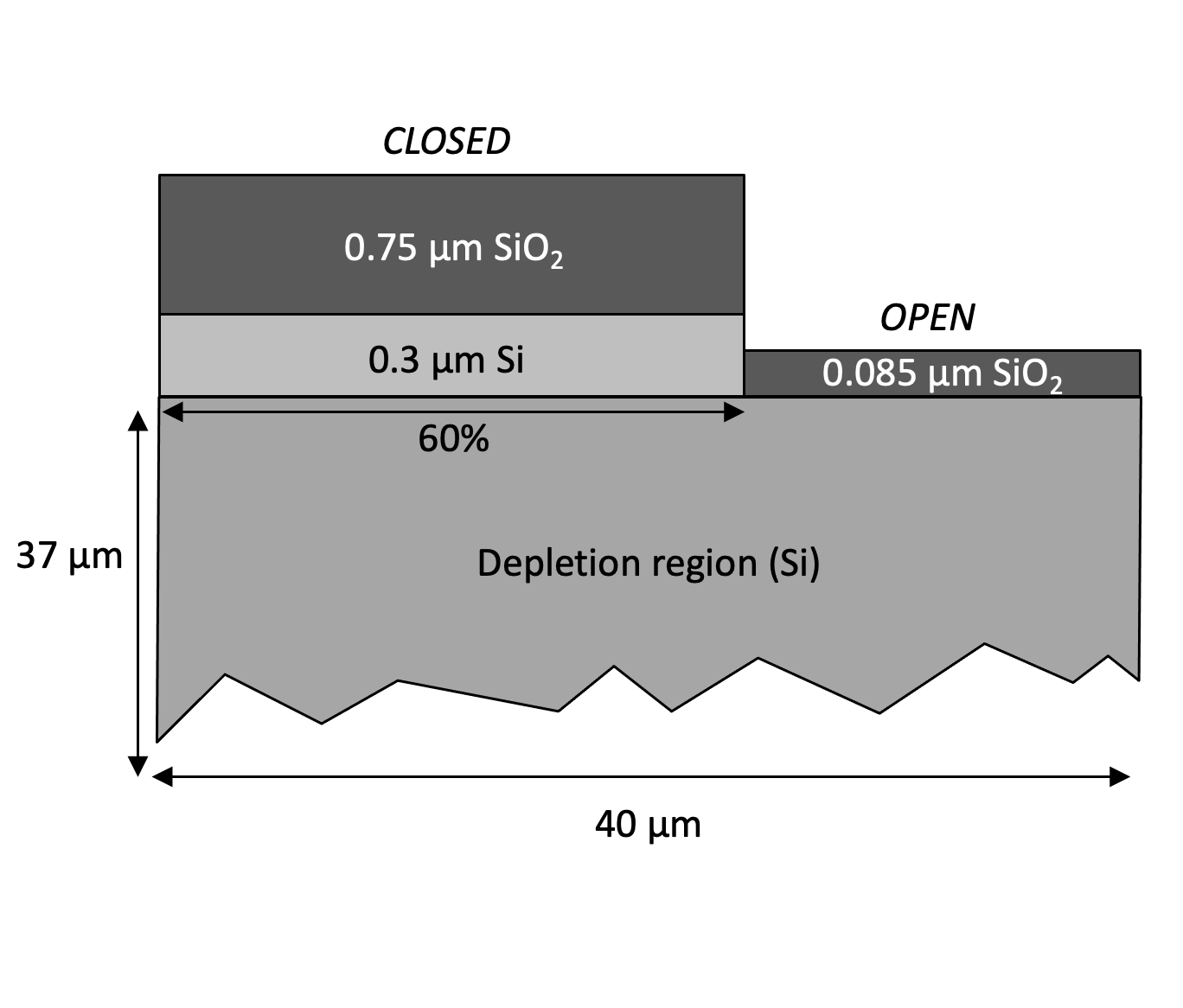}    
            }
      \caption{(left) Schematic view of the \textit{XMM-Newton} FPA geometry. (centre) Bottom view of the Geant4 mass model, with highlighted in yellow the MOS CCDs. (right) Schematic side view, not in scale, of the MOS pixel geometry implemented in the simulation, with the open and closed electrode structure on top of the depletion region covering 60\% and 40\% of the pixel area respectively. }
         \label{fig:fpa}
   \end{figure*}

We only implemented the forward section of the focal plane assembly (FPA), shared by the three EPIC cameras, containing the optical filters, the radiation shielding baffle, an Aluminium alloy baffle about 60 cm long placed at the top of the filter wheel, and surrounding structures modelled by an Aluminium\footnote{In the preliminary design published in \citet{2021SPIE11822E..1FF}, we wrongly used a Titanium proton shield as reported in \citet{nar01} but not confirmed in the EPIC official literature.} forward proton shield, 100 mm long, labelled as a proton shield in Fig. \ref{fig:fpa} (left panel). 
While the presence of the radiation shielding is fundamental in reducing the non-focused particle-induced background (for example induced by galactic cosmic rays), in case of protons entering the field of view the secondary scattering at the shield inner surface can potentially increase the flux at the detector.
The EPIC door and the filter wheel were not included. Instead, we connected the two proton shields with an additional Aluminium truncated cone. 
The filter wheel is equipped with three different optical blocking filters \citep{Struder2001} to reduce contamination from infrared, visible and UV light collected in the field of view. With a diameter of 76 mm and placed at 10 cm from the focal plane, the observer can select among two thin filters, one medium filter and one thick filter. Their configuration is reported in Table \ref{tab:filt}, with Polyimide and Polypropylene abbreviated as Pl and PP, respectively. 
\begin{table}[!h]
\begin{center}
\caption{\label{tab:filt}Composition and thickness of \textit{XMM-Newton} optical filters.}
\begin{tabular}{c|c|c|c}
\hline
\hline
\multirow{2}{*}{Filter} & \multirow{1}{*}{layer 1} & \multirow{1}{*}{layer 2} & \multirow{1}{*}{layer 3}\\
&[$\mu$m] & [$\mu$m]& [$\mu$m]\\
\hline
Thin  & Al (0.04)& Pl, (0.16)&\\
Medium  & Al (0.08)& Pl, (0.16)&\\
Thick  & Sn (0.045)& Al (0.11)& PP (0.33)\\
\hline
\end{tabular}
\tablefoot{The layer numbering starts at the mirror side.}
\end{center}
\end{table} 
\\
The MOS camera is composed by seven front-illuminated CCDs operating from 0.5 to 12 keV. Each CCD is divided into $600\times600$ pixels, each with an area of $40\times40$ $\mu$m$^{2}$. The central CCD is at the focal point on the optical axis of the telescope, while the others are at a distance of 4.5 mm towards the mirror, to approximately reproduce the focal plane curvature. Each CCD has a 300 $\mu$m-wide dead region on three sides. They are rotated and disposed in order to cover the dead sides as much as possible, to maximise the exposed area. An Aluminium ring, 100 $\mu$m thick, is placed on top the CCDs to represent the camera metalwork limiting the field of view to 62 mm of diameter. The Geant4 model of the MOS CCDs is shown in Fig. \ref{fig:fpa} (central panel).
   \begin{figure}
   \resizebox{\hsize}{!}
            {\includegraphics{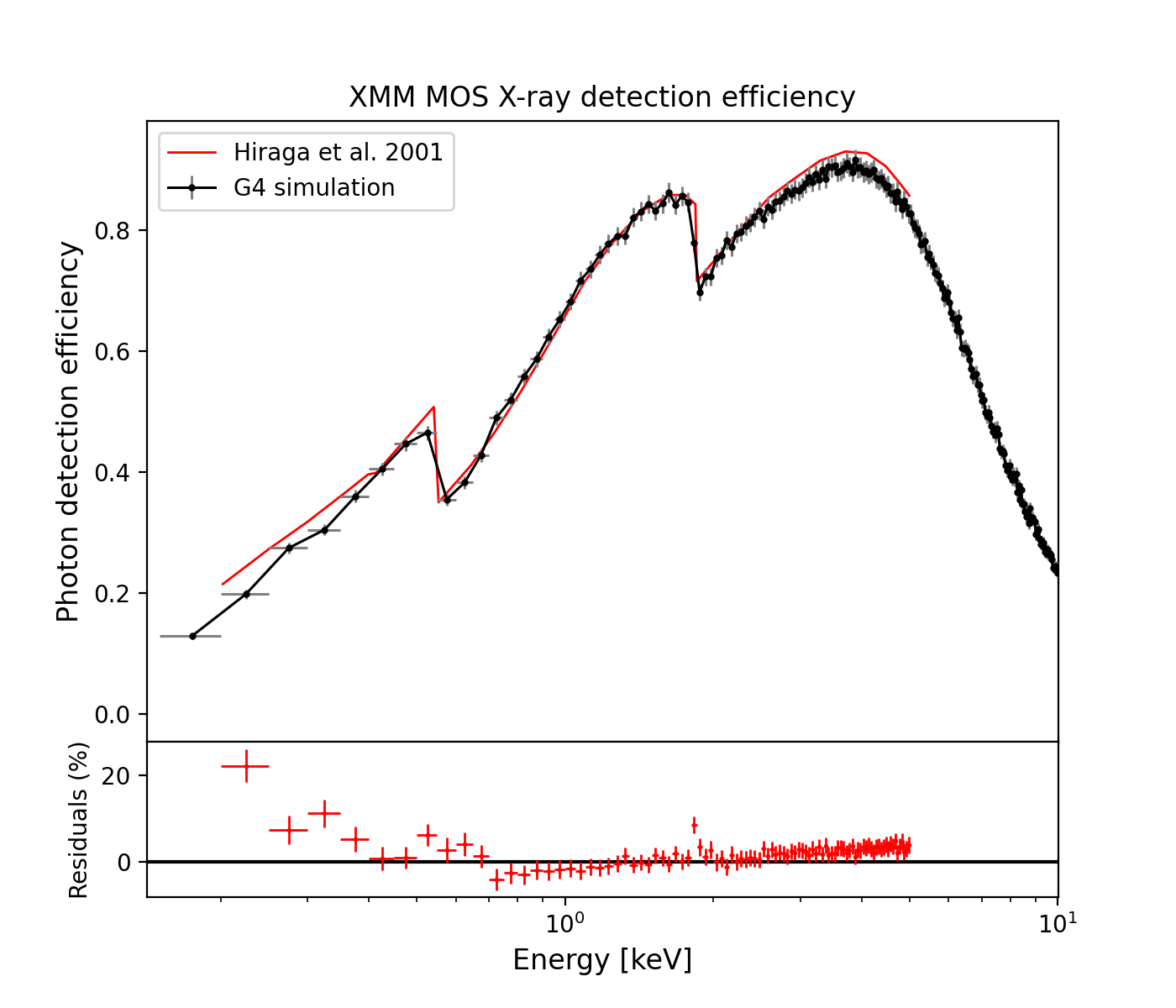} 
            }
      \caption{Comparison of the measured (red line) and simulated (black line) MOS X-ray detection efficiency (the error bars refer to the simulation statistical uncertainty). }
         \label{fig:eff}
   \end{figure}
\\
The MOS conventional 3-phase front-illumination device is characterised by an open electrode structure \citep{1996SPIE.2808..414H} where one of the electrodes was partially etched, that is, holes were cut through it, to increase the X-ray detection efficiency at low energies. The resulting device structure \citep{Hiraga2000, 2014MNRAS.445.2146F} is divided into a 60\% portion covered by the standard closed electrode structure, composed by a layer of Si and SiO2 , and a 40\% area of open electrode with only SiO2. The depletion region, made of Silicon, is 37 $\mu$m thick. Fig. \ref{fig:fpa} (right panel) shows a not in scale schema of the pixel geometry implemented in Geant4. The X-ray detection efficiency generated by the MOS CCD pixels was measured on-ground by sampling different positions of the array \citep{Hiraga2000}. The resulting mean pixel X-ray response is compared in Fig. \ref{fig:eff} to the Geant4 simulation of the MOS quantum efficiency obtained by illuminating the array with a uniform photon beam from 0.2 to 10 keV. The two curves are in very good agreement above 0.5 keV, with residuals within some percentage. Towards lower energies we find a degradation in the accuracy that reaches about 20\%. Since the work of \citet{Hiraga2000} does not include uncertainties in the curve, it is difficult to explain the source of the discrepancy. However, the general agreement well within 20\% positively verifies the Geant4 implementation of the MOS camera, a key factor in correctly estimating the proton energy losses.
\\
The pn camera is instead back-illuminated, composed of four quadrants each having three CCDs with $200\times64$ pixels, with a pixel size of $150\times150$ $\mu$m$^{2}$ and a total imaging area of $6\times6$ cm. The fully depleted Silicon thickness is 300 $\mu$m. Being back-illuminated, no read-out devices are present in front of the pn camera, simplifying its Geant4 implementation. In addition, no metalwork is placed on top of the camera and contrary to the MOS, the CCD area is fully exposed to the soft proton flux.
   \begin{figure}
   \resizebox{\hsize}{!}
            {\includegraphics{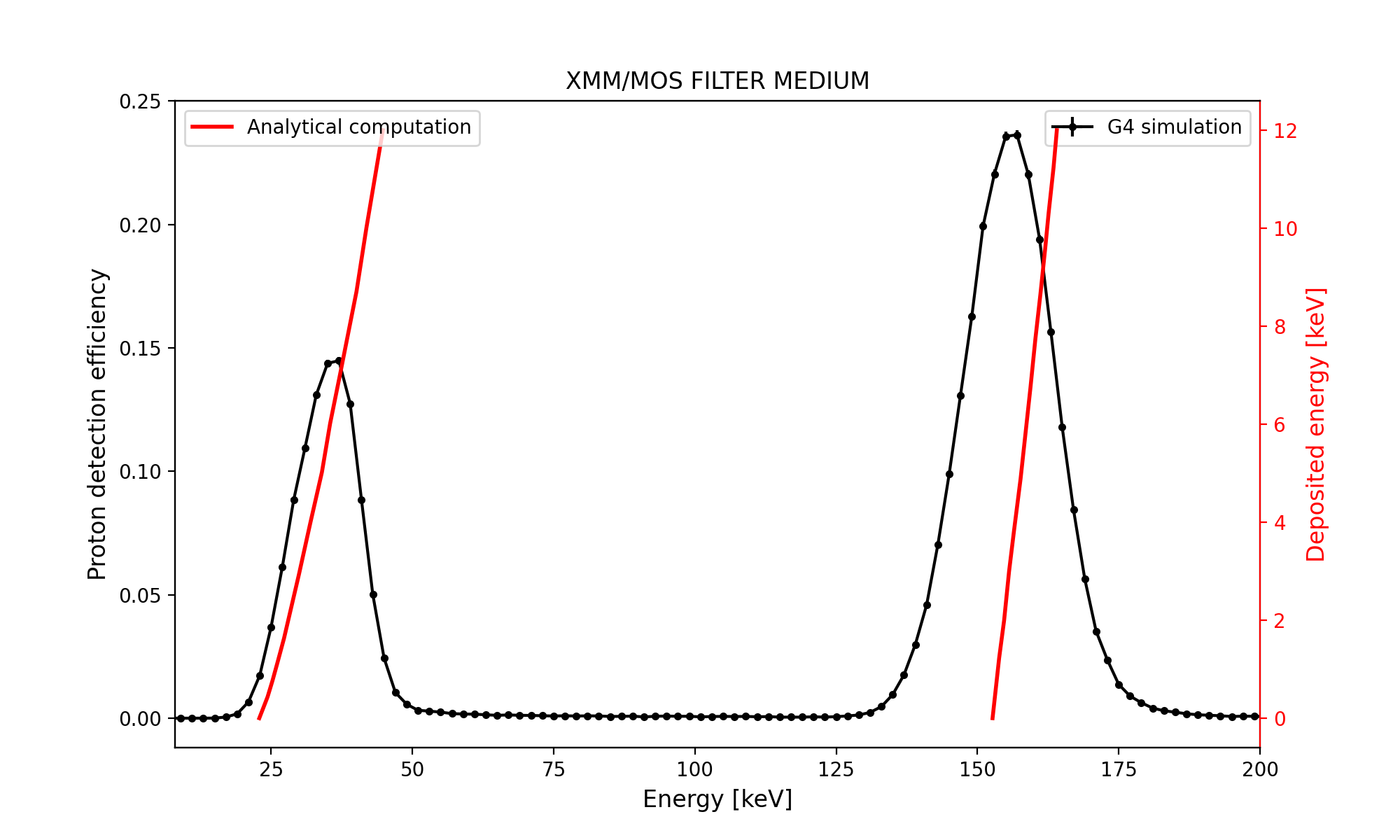} 
            }
      \caption{Simulated, in black, proton transmission efficiency in the 8 – 200 keV energy range for the MOS and medium filter. The analytical computation based on tabulated NIST proton stopping power predicts the proton energy range for which the MOS is likely to register a count, with an energy labelled by the red y-axis. }
         \label{fig:trans}
   \end{figure}
      \begin{figure*}
   \resizebox{\hsize}{!}
            {\includegraphics[width=0.35\linewidth]{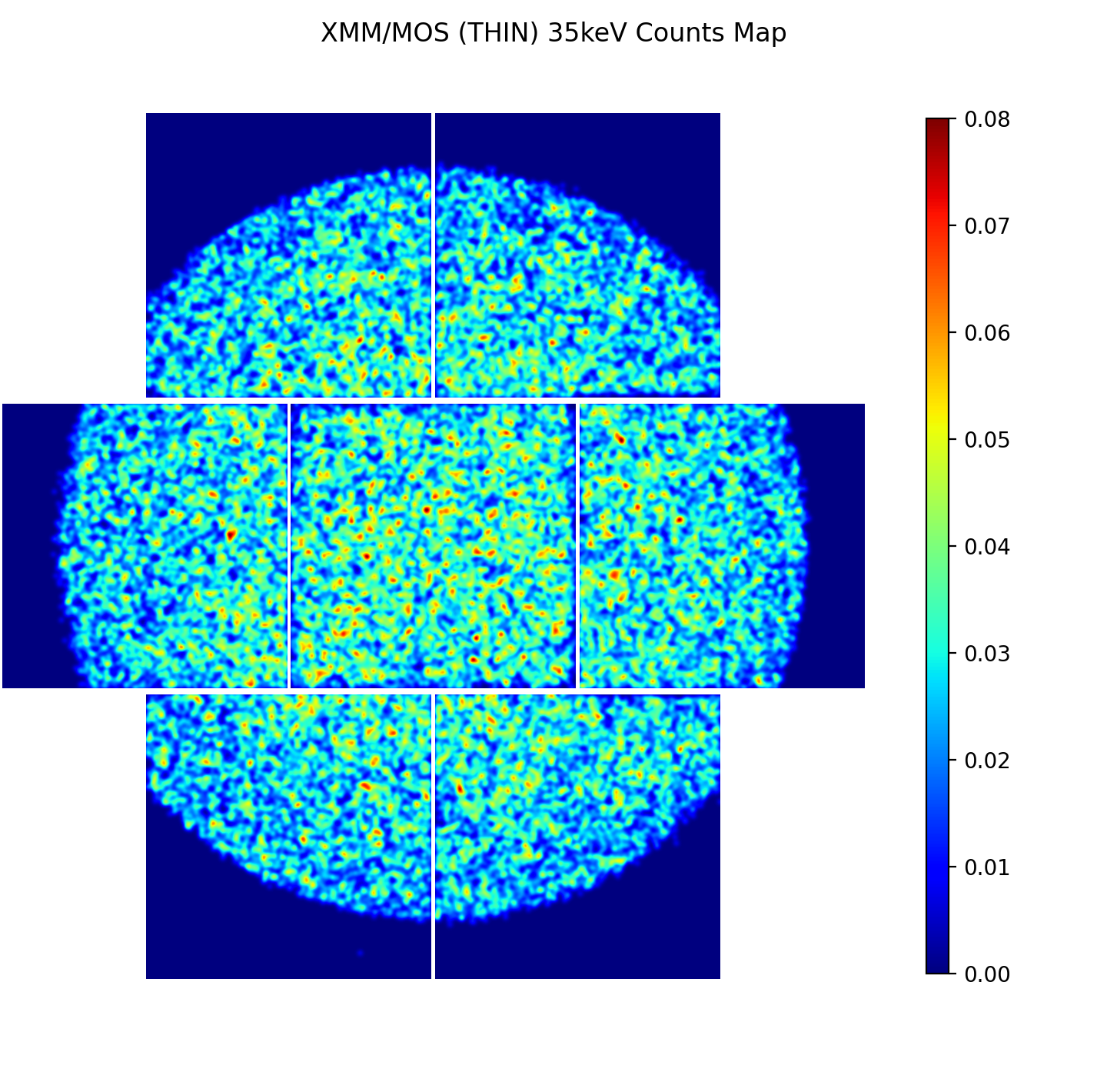}
            \includegraphics[width=0.5\linewidth]{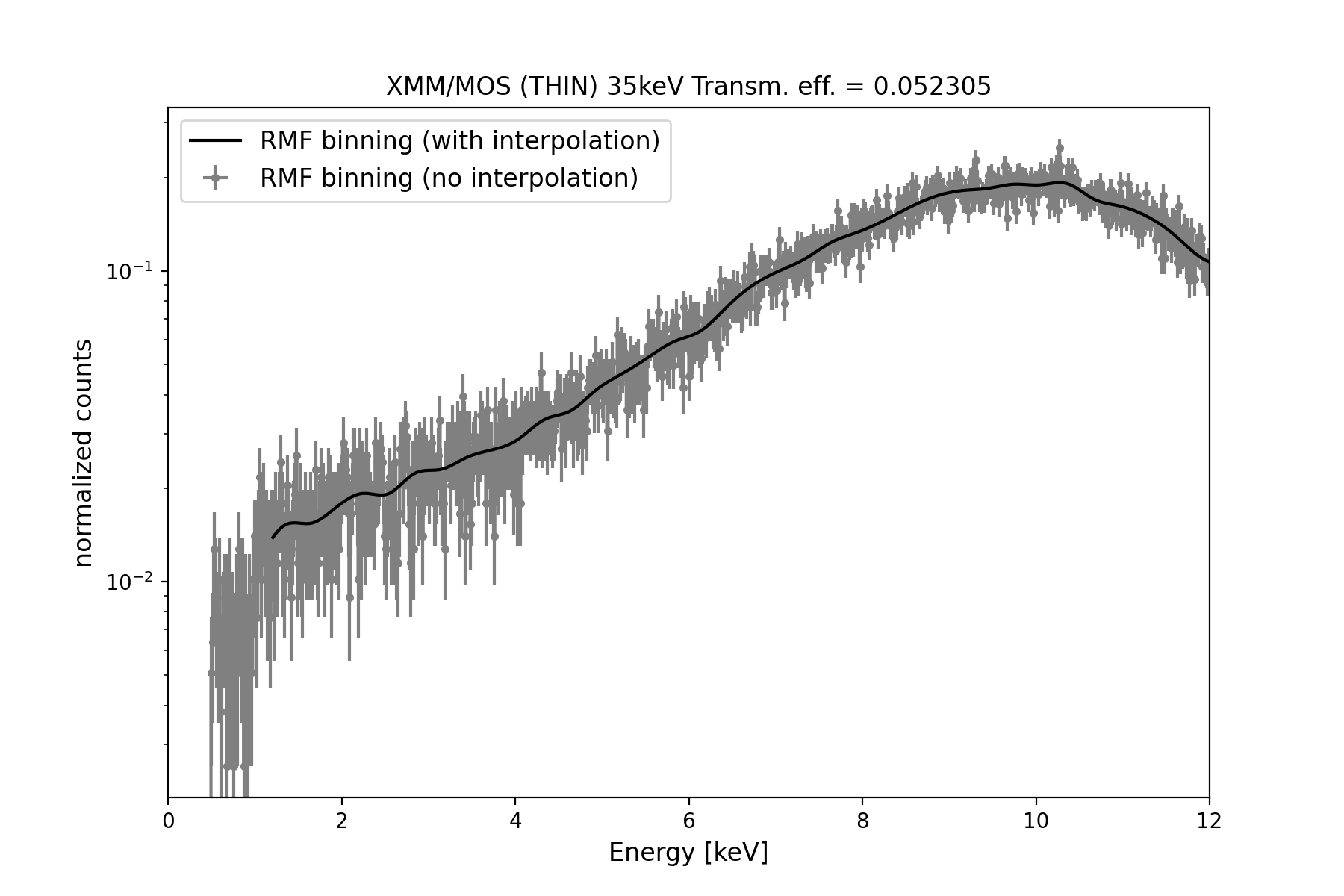}}
            \\
     \resizebox{\hsize}{!}
            {\includegraphics[width=0.42\linewidth]{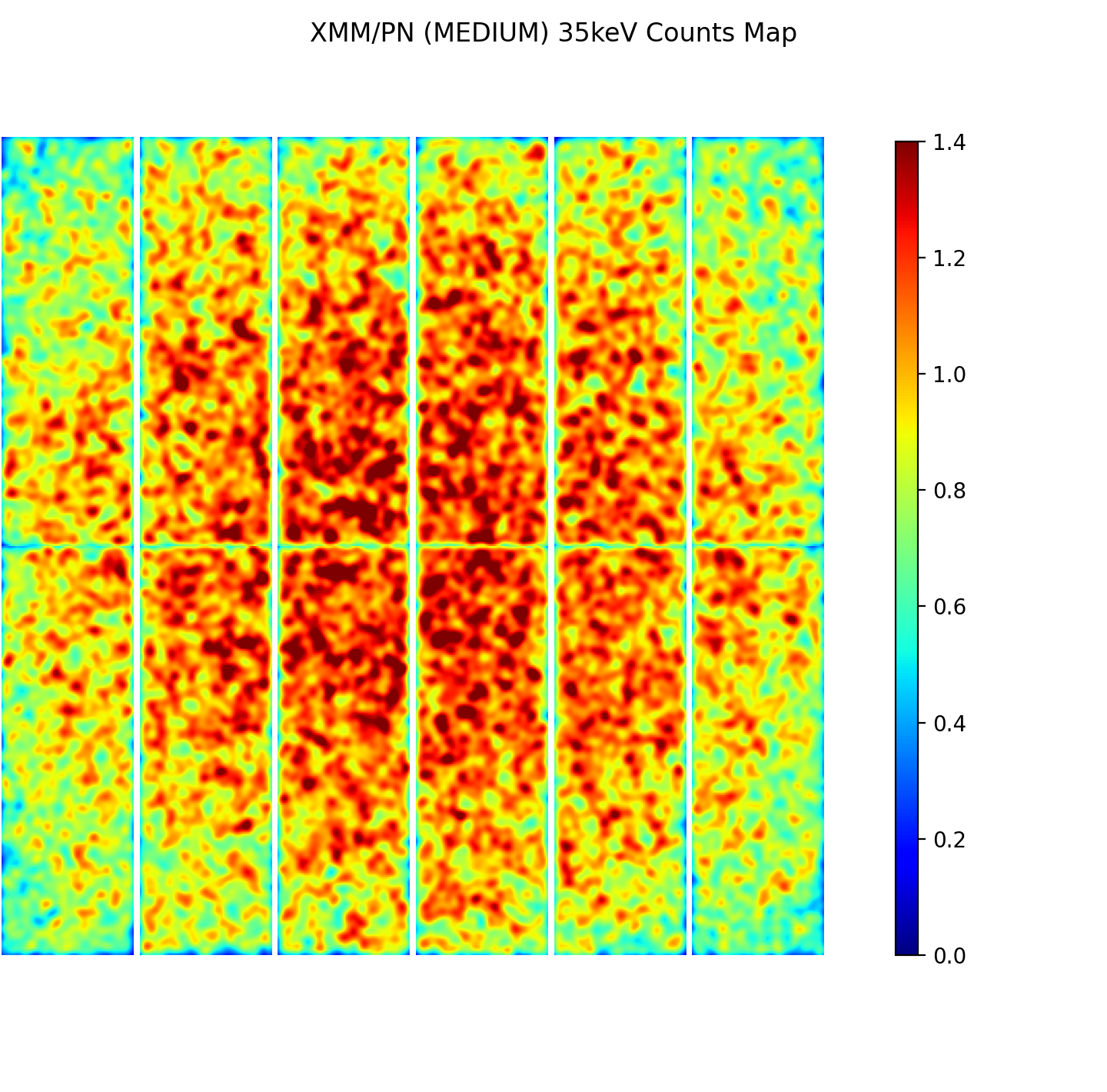}    
            \includegraphics[width=0.58\linewidth]{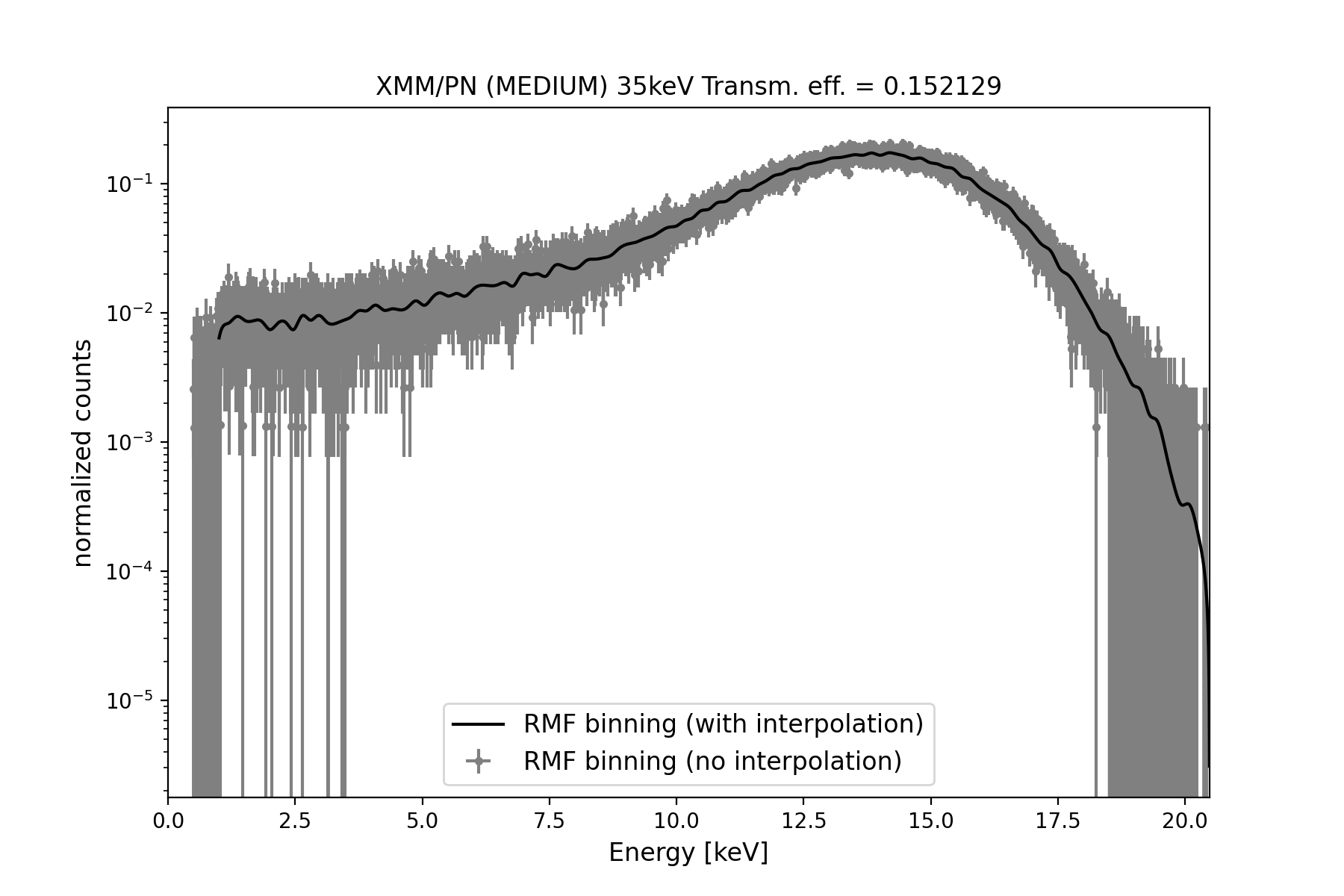}    
            }
      \caption{Count map (left) and interpolation of the detected spectral distribution (right) for an input energy of 35 keV. The top and bottom panels refer to the MOS (thin filter) and pn (medium filter) simulations.}
         \label{fig:maps}
   \end{figure*}
\subsection{FPA transmission efficiency}\label{sec:trans}
While soft X-ray photons are either absorbed by the optical filters or pass through with unaltered energy, protons continuously lose their energy along the path. For this reason, the filter configuration changes both the transmission efficiency, included in the effective area, and the energy redistribution, changing the spectrum of the proton flares. The same applies to the crossing of the MOS electrode structure, that is not present in the pn CCDs.
\\
Soft protons mainly interact by means of ionisation and scattering when crossing the optical filters and the MOS electrodes.
The precision of the Bethe-Bloch formula, describing the continuous energy loss due to ionisation, degrades below 2 MeV and Geant4 interpolates instead the tabulated National Institute of Standards and Technology (NIST) proton stopping power \citep{2017ExA...tmp...50l}, which below 500 keV is based on experimental data. The declared uncertainties\footnote{https://physics.nist.gov/PhysRefData/Star/Text/programs.html} range from 5\% to 10\% at 100 keV, 10\% to 15\% at 10 keV, and at least 20\% to 30\% at 1 keV.
The limitation in the transmission simulation accuracy is coupled with the intrinsic statistical nature of the ionisation process that causes large fluctuations of the energy losses through thin ($<1$ mm) layers such as the optical filters and MOS electrodes.
\\
Laboratory measurements of the proton transmission would estimate the level of systematic uncertainty introduced by the Geant4 simulations and provide an overall validation of the FPA effect in the proton detection. Lacking such measurements, we can only compare the efficiency with the one obtained using the NIST tabulated stopping power, which are averaged over the total thickness and do not include the angular spread that changes the mean free path through the layers. 
The transmission efficiency is defined as the ratio of the protons depositing energy in the EPIC operative band over the number of protons entering the baffle field of view. The etching of the MOS electrodes causes two peaks in the MOS efficiency (Fig. \ref{fig:trans}), one centred at $\sim35$ keV if protons cross the open electrode, and one centred at $\sim160$ keV if they encounter instead the thicker closed electrode.
The analytical computation (red lines of Fig. \ref{fig:trans}) defines an energy range where protons are able, according to the tabulated stopping power, to reach the MOS pixels with an energy $<$12 keV. The resulting ranges are in general agreement with the simulated peaks. The full characterisation of the energy losses and angular spread through the layers was reported in \citet{2021SPIE11822E..1FF}.

\subsection{Energy redistribution}\label{sec:ene}
Mono-energetic proton beams from 2 to 300 keV, with a bin of 1 keV, are isotropically simulated at the mirror entrance and their energy, position, and direction after scattering is extracted at the entrance of the proton shield baffle. 
   \begin{figure*}
   \resizebox{\hsize}{!}
               {\includegraphics[width=0.35\linewidth]{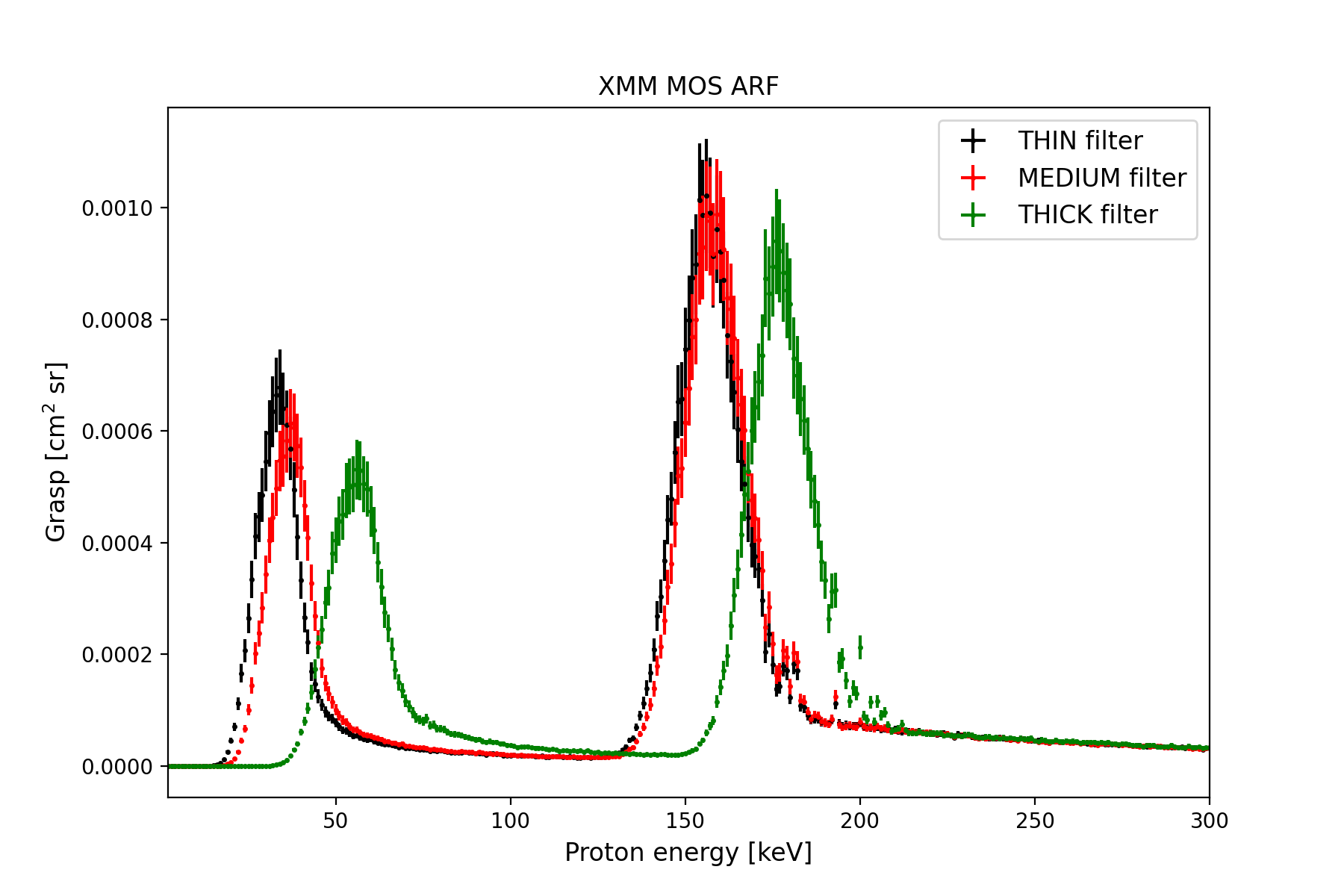}
            {\includegraphics[width=0.35\linewidth]{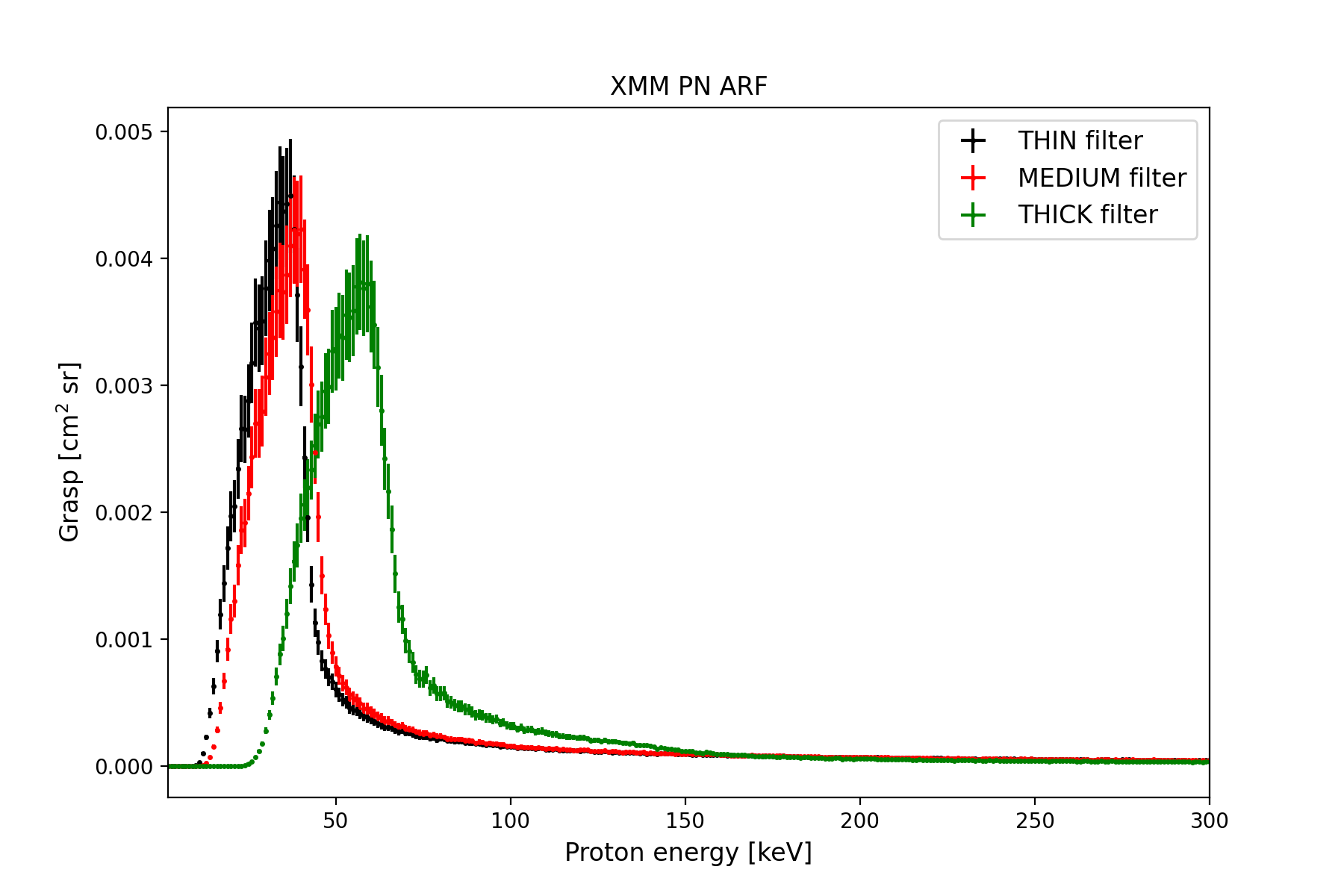}}
            }\\
            {\includegraphics[width=0.5\linewidth]{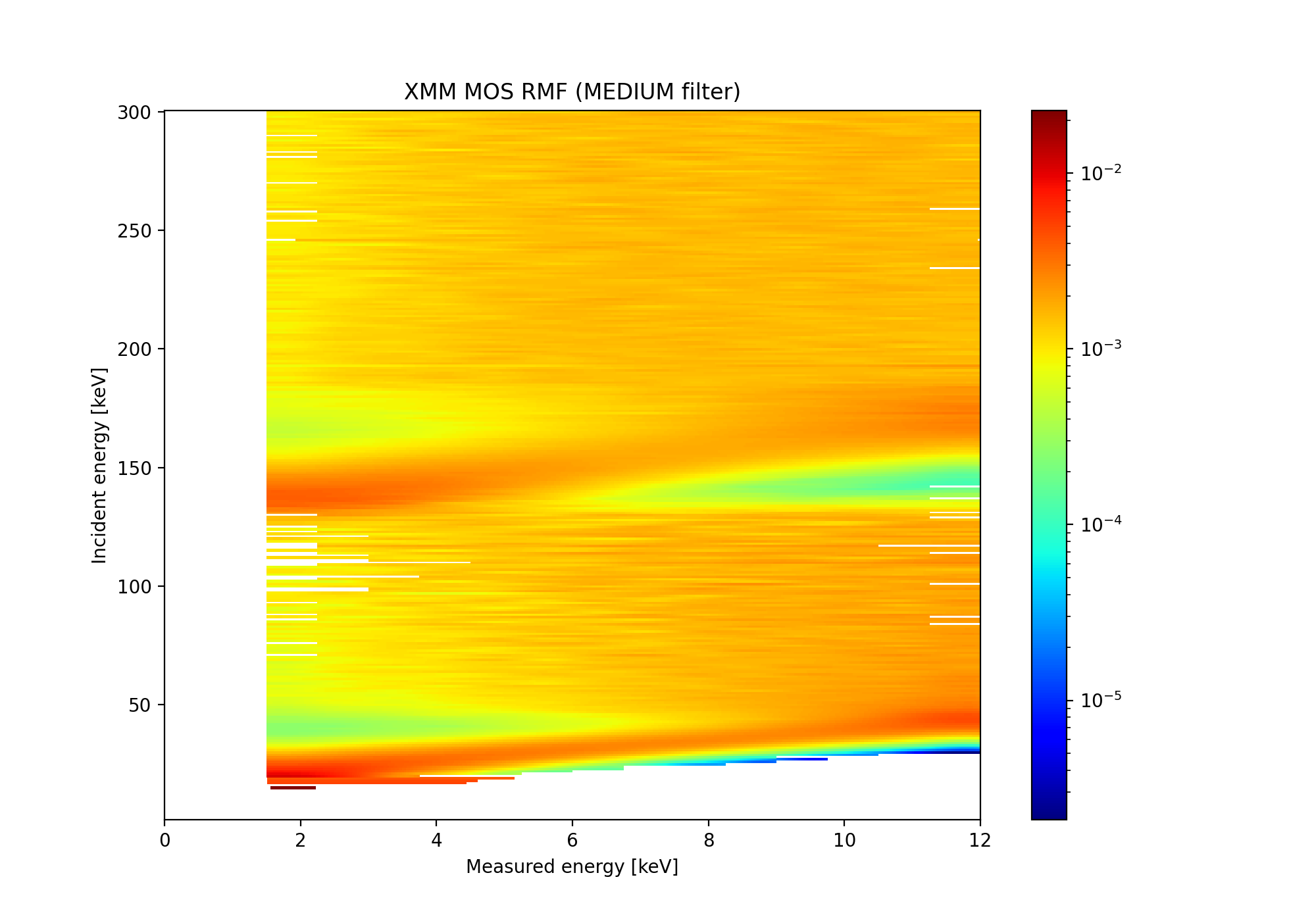}
            {\includegraphics[width=0.5\linewidth]{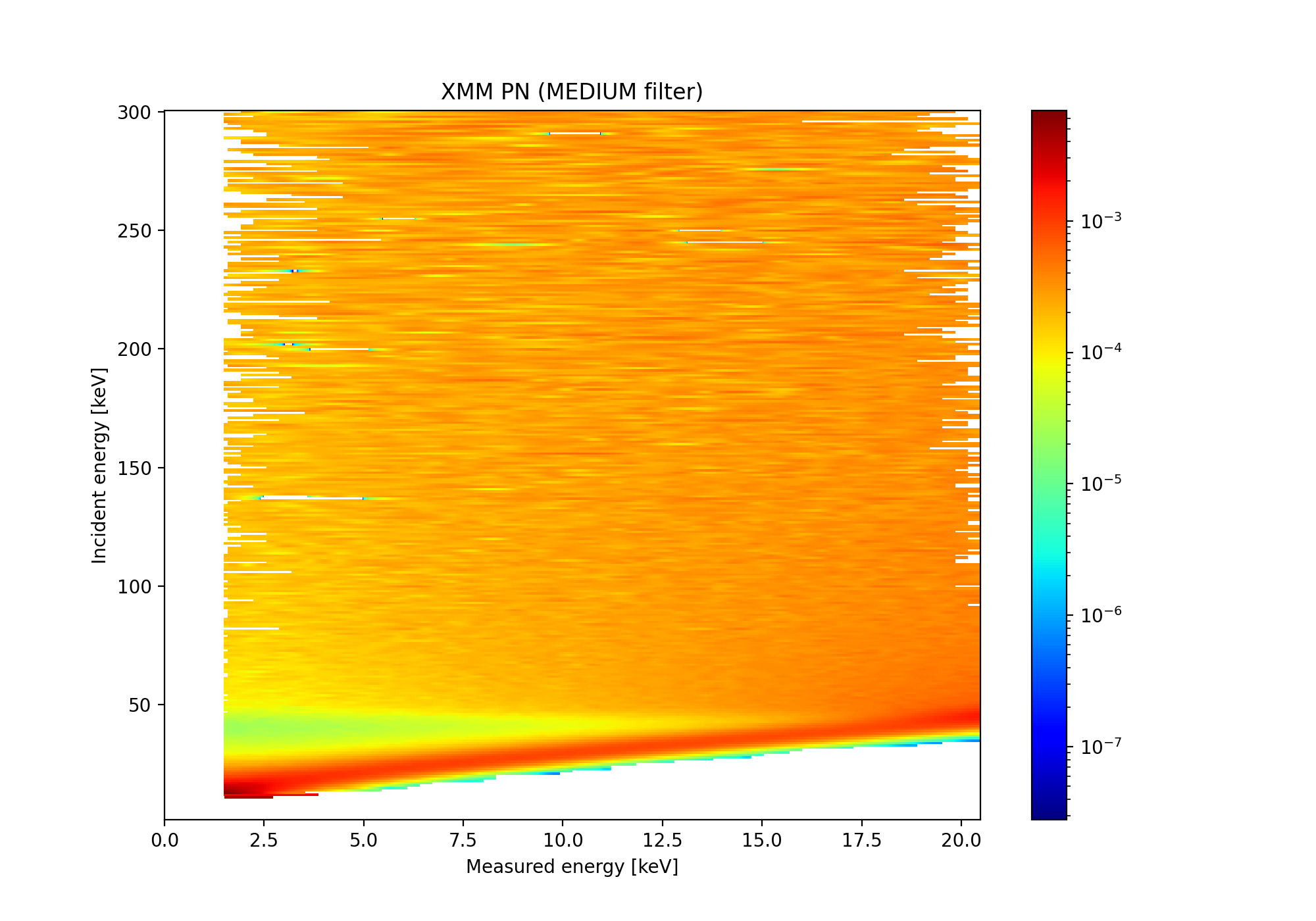}}
            }
      \caption{Grasp curves (top) stored in the ARF and RMF (bottom) for the MOS (left) and pn (right) CCDs. A 10\% systematic uncertainty has been added to the grasp curves.}
         \label{fig:rmf}
   \end{figure*}
For each input energy, the detected energy spreads through the full operative range of the detectors and in the full area exposed to the field of view, as shown by the simulated counts maps of Fig. \ref{fig:maps} (left panels). 
The spectra are first binned using the X-ray RMF channels (see Sect. \ref{sec:resp}), as shown in grey in Fig. \ref{fig:maps} (right panels). Statistical errors dominate the plot, and the energy distribution cannot be directly stored in the proton RMF. 
The adopted solution was to rebin and interpolate the histogram, the black curve of Fig. \ref{fig:maps} (right panels), and use the interpolation function to fill the RMF. If the number of counts was not enough to produce a histogram in the first place, a constant was used along the covered energy range. This method allowed us to automatically generate the RMF model for all the input energies despite the limited statistics, but it also lead to some caveats in the proton response usage. The interpolation function produces artefacts at the border of the energy range, in particular when the count number is low. We compared (see Sect. \ref{sec:ver}) simulations using the response matrix against a standard simulation - where no artefacts are present since there is no interpolation - and defined a validity energy range for the use of the response files. 
We also note the presence in the spectral redistribution of a cut at about 1 keV. Since the accuracy of the Geant4 modelling of the proton stopping power degrades below 1 keV, the simulated proton efficiency at the detector also features a reduction at low-energy that is not physical but caused by internal limits in the Geant4 physics model. For this reason, the energy redistribution is modelled from 1.5 keV to the upper threshold of the instrument X-ray RMF.

\section{Building the proton response matrix}\label{sec:build}
The energy distribution is binned according to the respective instrument X-ray channels. For the MOS, a total of 800 channels, an energy width of 15 eV, and energy boundaries of 0 – 12 keV, for the PN a total of 4096 channels, a mean energy width of 5 eV, and energy boundaries of 0 - 20 keV.
\\
From the full simulation pipeline, we obtain for each input energy a list of counts with energy, position and pattern flag. We compute then by interpolation the energy probability distribution in the instrument channels, normalised to 1. 
The effective area is the product of the mirror grasp and the transmission efficiency, computed as the fraction of protons that deposit an energy amount within the detector working range. The RGS is not present in the Geant4 mass model, but its effect is included as a reduction of 50\% of the MOS grasp. This assumption is confirmed by the simulations performed by \citet{nar02b} to investigate potential radiation damage from soft proton propagation on \textit{XMM-Newton} and \textit{Chandra}.
A different proton response was produced for each combination of optical filter type, and focal plane instrument. The pattern analysis shows that 99.9\% of the simulated counts has a pattern of 0, that is, only one pixel triggered for each event, similarly to what observed in the in-flight focused non X-ray background. For this reason, all the proton response files were produced selecting only singles (\textit{PATTERN = 0} for the EPIC cameras). 
At first, we planned to connect the FPA to the mirror simulation by using the same proton list at the mirror output as input for the FPA simulation, to reproduce the exact angular, spatial, and energy distribution, but the computational time required to achieve a minimum statistical level to model the energy distribution turned out to be unfeasible for the available CPU resources. The solution was to model the angular and energy distribution of the protons at the entrance of the proton baffle and then exploit the Geant4 Monte Carlo generator to randomly sample the protons within the modelled distribution. 
This approach averages the spatial distribution of the protons, removing the vignetting effect of the mirror \citep{2016SPIE.9905E..6WF}, and assumes a uniform angular and energy distribution at the baffle entrance, with no dependence of the proton energy on its direction.
Since the response files are produced for full field of view observations, averaging the vignetting effect doesn't affect their accuracy. We tested the modelled distributions by comparing the simulated spectra obtained with the response files with standard end-to-end simulations using the exact proton list at the mirror exit. The result is presented in Sect. \ref{sec:vv}
\\
The ARF curves for the thin, medium, and thick filters are shown in Fig. \ref{fig:rmf} (top panels), with an applied 10\% systematic uncertainty derived from the Geant4 accuracy in simulating the proton transmission below 100 keV (Sect. \ref{sec:trans}). The MOS open and closed electrode structure causes two peaks in the grasp, with protons losing additional $\sim100$ keV for just hitting different regions of the CCDs. A long tails extends to 300 keV and beyond because of secondary scattering with the baffle. The tail, because of the power-law nature of the proton input spectral distribution, does not impact the total grasp. 
\\
The resulting energy redistribution matrix of the MOS and pn CCDs, plotted as incident vs measured energy and stored in the RMF files, is shown in Fig. \ref{fig:rmf} (bottom panels) for the medium filter. For each input proton energy, the RMF stores the energy distribution detected by the instruments normalised to 1. 
The effects of the transmission of the two regions of the electrodes are also clearly visible in the RMF. The RMF and ARF files are averaged on the camera field of views.
\\
The validity energy ranges, required to avoid artefacts by the interpolation model and defined by the comparison with standard simulations (see Sect \ref{sec:vv}) are 2 -- 11.5 keV for the MOS and 2 -- 19.5 keV for the pn. The user should ignore the channels outside the validity ranges when performing their analysis.
Considering the 15-20\% systematic uncertainty in the mirror geometry model and the systematic error (10\%) introduced by the Geant4 proton transmission models, we estimate a total uncertainty of 30\% to the simulations obtained with the proton response files.

\section{Verification and validation results}\label{sec:vv}
The verification of the proton response files consisted in checking their technical correctness by simulating the MOS and pn X-ray spectrum with \textit{Xspec} using the response files, as a general user would do, and in parallel producing a spectrum with a standard simulation, where the protons are simulated at the mirror entrance with the same distribution used to generate the matrices and the background counts are collected at the detector. The validation process requires instead testing the response matrix with representative soft proton spectra, and comparing the best-fit models with proton space environment models extracted from independent measurements. An extensive test of the response files with in-flight proton flare observations is presented in Paper II.

\subsection{Proton radiation models}\label{sec:input}
   \begin{figure}
   \resizebox{\hsize}{!}
            {\includegraphics[width=0.35\linewidth]{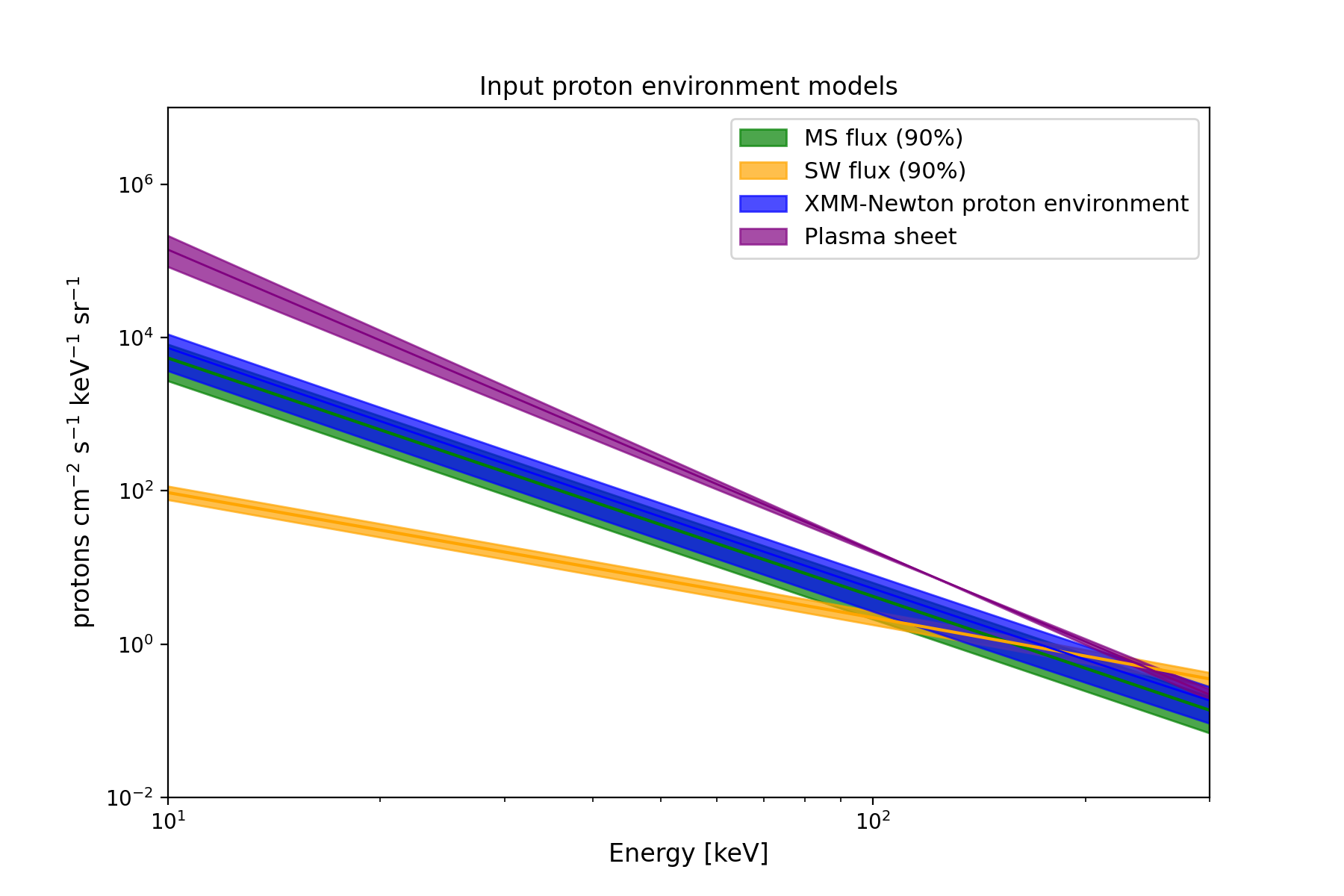}}
      \caption{Magnetosheath (MS), solar wind (SW), plasma sheet, and \textit{XMM-Newton} orbital proton spectral models (maximum cumulative fraction of 90\%) used as input in the verification and validation activity.}
         \label{fig:input}
   \end{figure}

The EXTraS\footnote{http://www.extras-fp7.eu} (Exploring the X-ray Transient and variable Sky) project \citep{2017ExA....44..297M}, funded by the FP7 European programme, provided an unbiased database of \textit{XMM-Newton} EPIC blank sky observations from 2000 to 2012. This unprecedented collection of data was used in the ESA AREMBES project \citep{AREMBES} to characterise both the unfocused and focused observed particle background. From the work of \citet{2017ExA....44..273G}, 
\textit{XMM-Newton} during this period spent most of the time south of the ecliptic plane.
Depending on the season, the apogee exited the bow shock, the intersection between the solar wind and the Earth's magnetic field, towards the Sun and fully exposed to the solar wind or it stayed completely inside the magnetosphere, where in the anti-Sun forms a long tail of trapped particles, the magnetotail. The 49\% of EXTraS observations were spent in the magnetosheath, the plasma region behind the bow shock, hotter and more turbulent than the solar wind \citep{2013JGRA..118.4963D}, the 40.5\% in the magnetotail, only 4.2\% in the solar wind and 0.5\% in the plasma sheet, a sheet-like region in the equatorial magnetotail region. Moderate variations were found in the different regions, with the plasma sheet linked to the highest flaring mean rate.
\\
The spectral distribution of the proton component of the different regions of the magnetosphere was extracted from various in-flight measurements during the AREMBES project, to study the background of the \textit{Athena} mission initially planned in L2. We show in Fig. \ref{fig:input} the power-law models representative of the interplanetary solar wind (SW) and the magnetosheath (MS) extracted from \citep{Lotti2018}, the plasma sheet extracted from \citet{2018ApJ...867....9F}. They refer to an active state of the magnetosphere and the maximum flux encountered in 90\% of the operational time. The solar wind fluxes are about the same in L1 and L2 \citep{Lotti2018}, and we assume that the interplanetary environment predicted for \textit{Athena} is also valid along the \textit{XMM-Newton} orbit. For the magnetospheric environment, the \textit{Athena} models used the first two years of data from the GEOTAIL mission, which monitored on average farthest magnetotail regions (from 8 to 210 Earth radii) than the \textit{XMM-Newton} apogee. Without models for the \textit{XMM-Newton} proton environment, we use the models for the different regions produced by the \textit{Athena} study, but using the fraction of time spent in the different regions reported in \citet{2017ExA....44..273G} for \textit{XMM-Newton}.
For the magnetotail and magnetoplasma regions, for which we lack models, we assume the same model for the magnetosheath, but normalised for the different flaring mean rates. The result is reported in blue in Fig. \ref{fig:input}.
While the uncertainties for the magnetosphere models are based on the AREMBES reported uncertainties ($\sim50\%$), for the SW, for which more extended observations were available, we applied a 20\% uncertainty level. We must note that all models were produced from observations starting at about 50 keV (47 keV for the SW and 58 keV for the GEOTAIL data) and then extrapolated to lower energies. No actual measurements are available in the very soft proton regime.
\subsection{Technical verification}\label{sec:ver}
\begin{figure}
   \resizebox{\hsize}{!}
            {\includegraphics[width=1\linewidth]{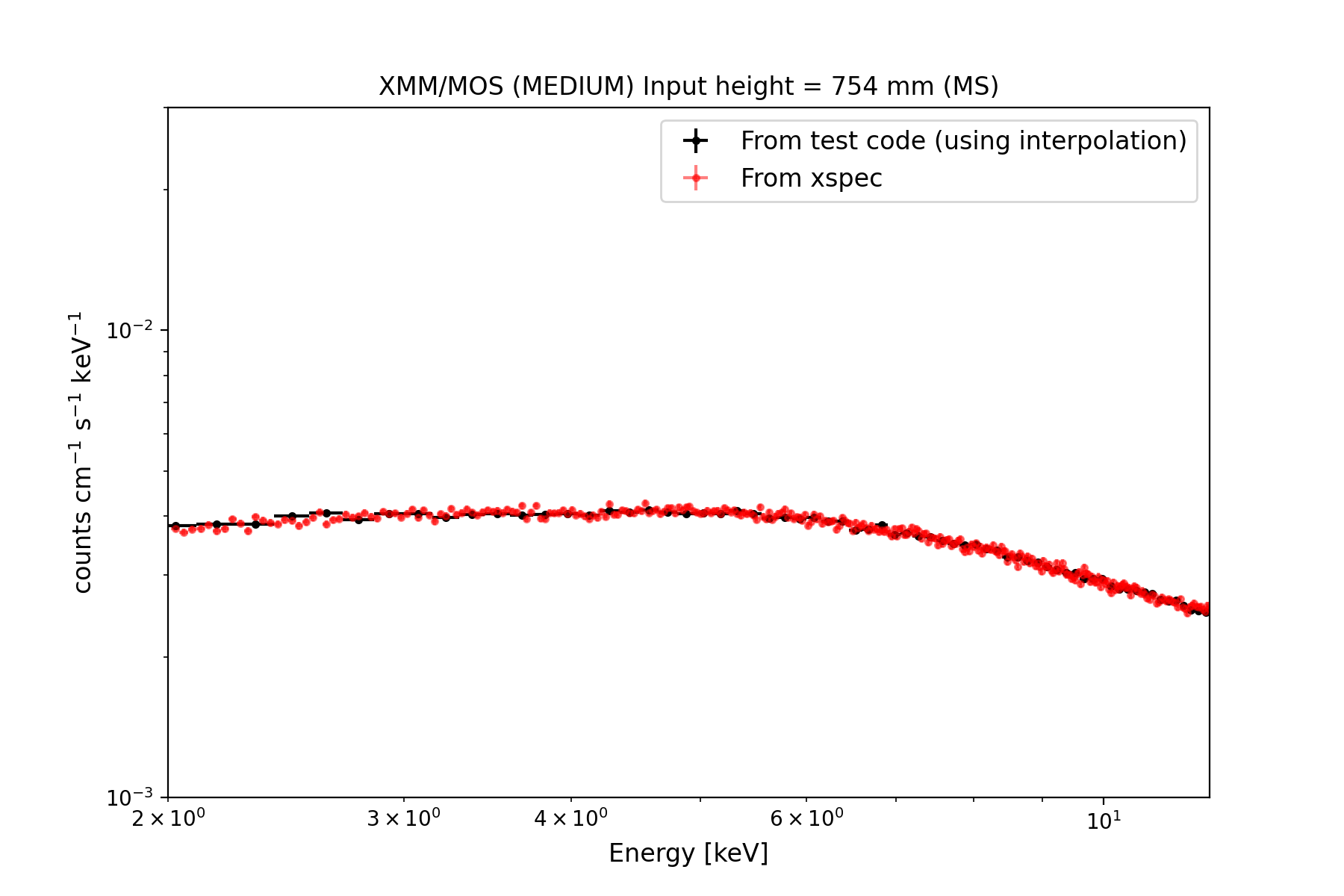} 
            }
            {\includegraphics[width=1\linewidth]{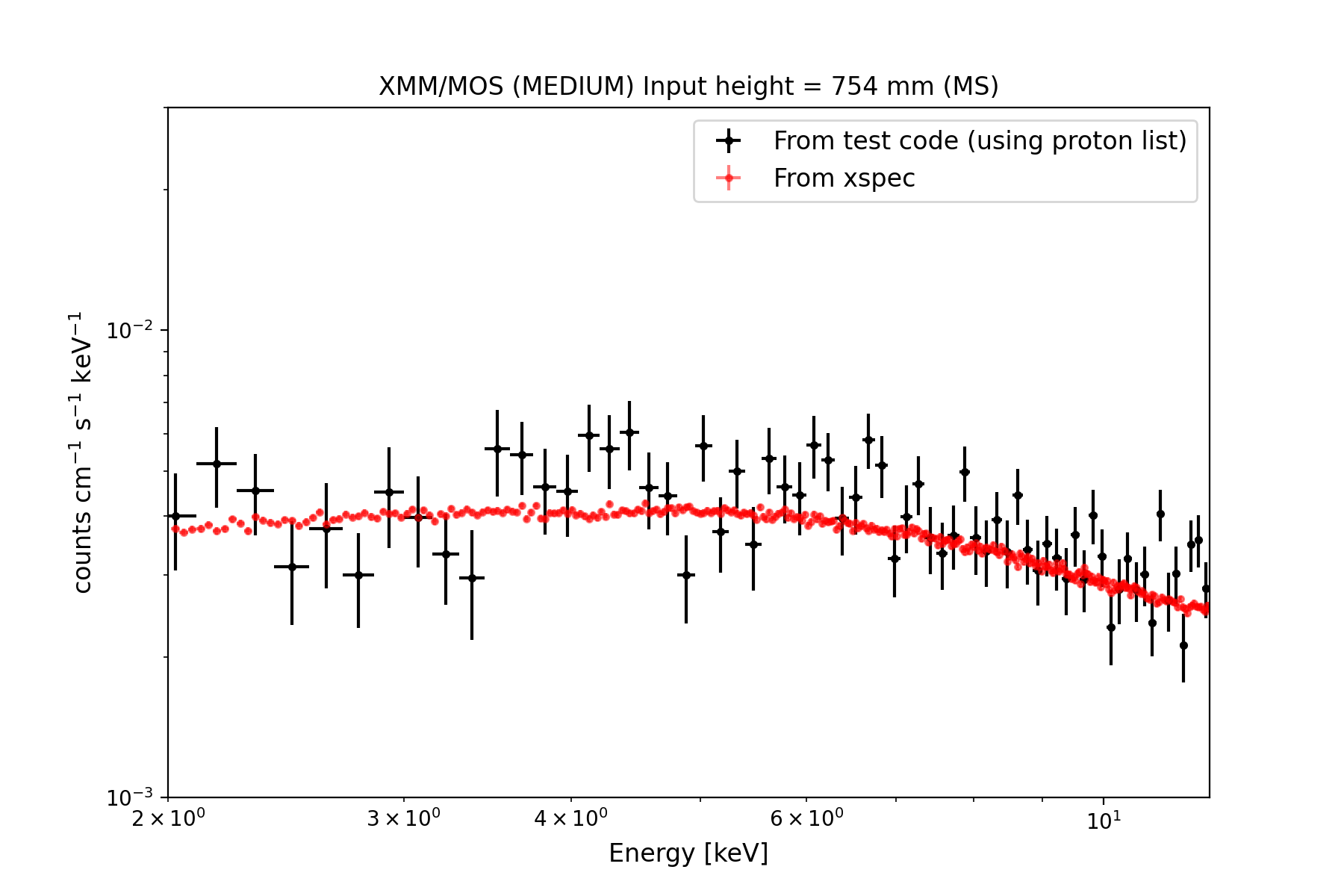}}
      \caption{Comparison of standard test simulations using modelled distributions (top) and exact proton lists (bottom) with simulations obtained in \textit{Xspec} using the proton response files. }
         \label{fig:plist}
   \end{figure}

In \textit{Xspec}, the simulated soft proton flux, in cts cm$^{-2}$ s$^{-1}$ keV$^{-1}$, is obtained by convolving the input model with the RMF and ARF files. 
We note that since the ARF stores the grasp of the telescope, the input model is normalised for the solid angle. 
The soft proton-induced background spectrum for the MOS (medium filter) is simulated in \textit{Xspec} for the input MS model and shown in Fig. \ref{fig:plist} (top panel, in red). The spectrum is compared to the standard Geant4 simulation (in black), where protons are propagated from the mirror to the focal plane. The same comparison is obtained in Fig. \ref{fig:plist} (bottom panel) against the Geant4 simulation using the exact proton list at the mirror exit and not the interpolation (see Sec. \ref{sec:build}). The resulting spectra are comparable, despite the limited statistics of the distribution using the proton list as input. These plots show that the response files correctly encode the Geant4 simulation.
   
\subsection{Validation with in-flight data}\label{sec:inflight}

The validation of the proton response files is only possible with the analysis of in-flight observations of soft proton flares, that can be analysed in \textit{Xspec} as an astrophysical observation. The best-fit model is then compared with our reference models for the proton environment (Fig. \ref{fig:input}), that are the maximum flux expected in 90\% of exposure time. Because of the flaring nature of soft proton flares, we can't use a random observation but we need to extract from archival studies an averaged spectrum representative of the maximum intensity in 90\% of the mission observation time. 
Such validation not only proves the accuracy of the response files but also allows us to study the origin of \textit{XMM-Newton} soft proton flares.

\subsubsection{Data preparation}
Soft proton flares are extremely unpredictable in duration, lasting from 100 s to hours, and intensity. We use the AREMBES's systematic analysis of the \textit{XMM-Newton} focused background \citep{2017ExA....44..309S} from the EXTraS archive of 13 years to extract the maximum soft proton intensity for a cumulative fraction of 90\%, that is, in 90\% of observation time. In this period  no significant variation in the soft proton transmission was detected in the MOS2 data \citep{2017ExA....44..273G}.
\\
The observations must be cleaned not only of source contamination, cosmic X-ray background (CXB), and instrumental noise, but also the level of the unfocused particle background. Since only the MOS detector provides an out-of-Field of view (out-of-FoV) region to measure the unfocused background component and because of the loss of the MOS1 CCD in 2005 for a micro-meteoroid impact, only EPIC MOS2 observations were used for this study. 
  \begin{figure}
   \resizebox{\hsize}{!}
            {\includegraphics[width=1\linewidth]{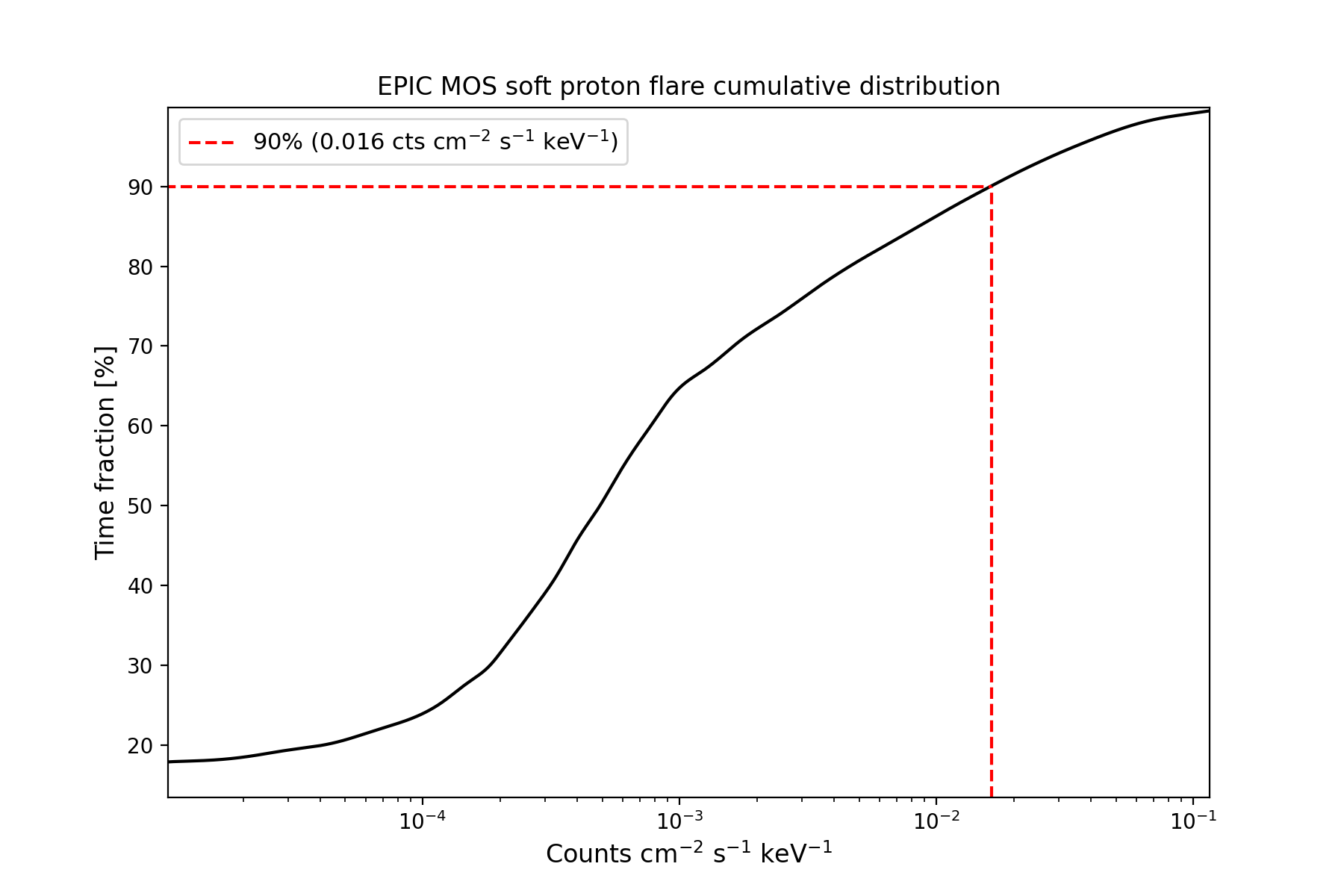} 
            }
      \caption{CDF of the EPIC MOS2 focused background count rate (medium filter). The dashed lines refer to the maximum rate expected in 90\% of exposure time. }
         \label{fig:cdf}
   \end{figure}
The focused background count rate was extracted with the following conditions:
\begin{itemize}
    \item only single and double events and standard flags to avoid bright columns and pixels; 
    \item energy selection in the 7 -- 12 keV band to avoid CXB contamination;
    \item exclusion of the 9.4 -- 10 and 11 -- 12 energy bands to avoid the Gold fluorescence line;
    \item only exposures in the Full Window mode to obtain a uniform data-set.
\end{itemize}
 \begin{figure}
  \centering
            {\includegraphics[width=0.9\hsize]{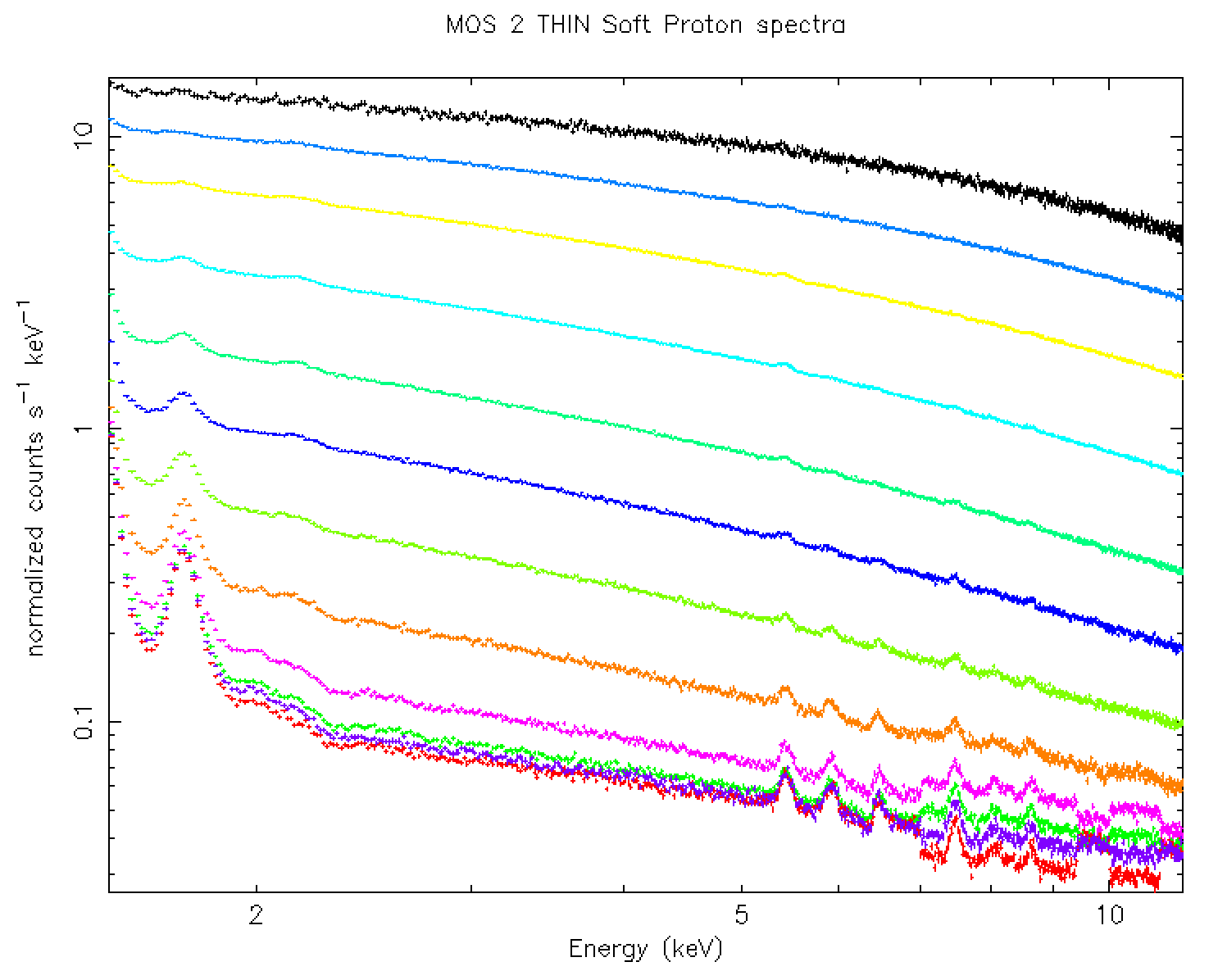}}
            {\includegraphics[width=0.9\hsize]{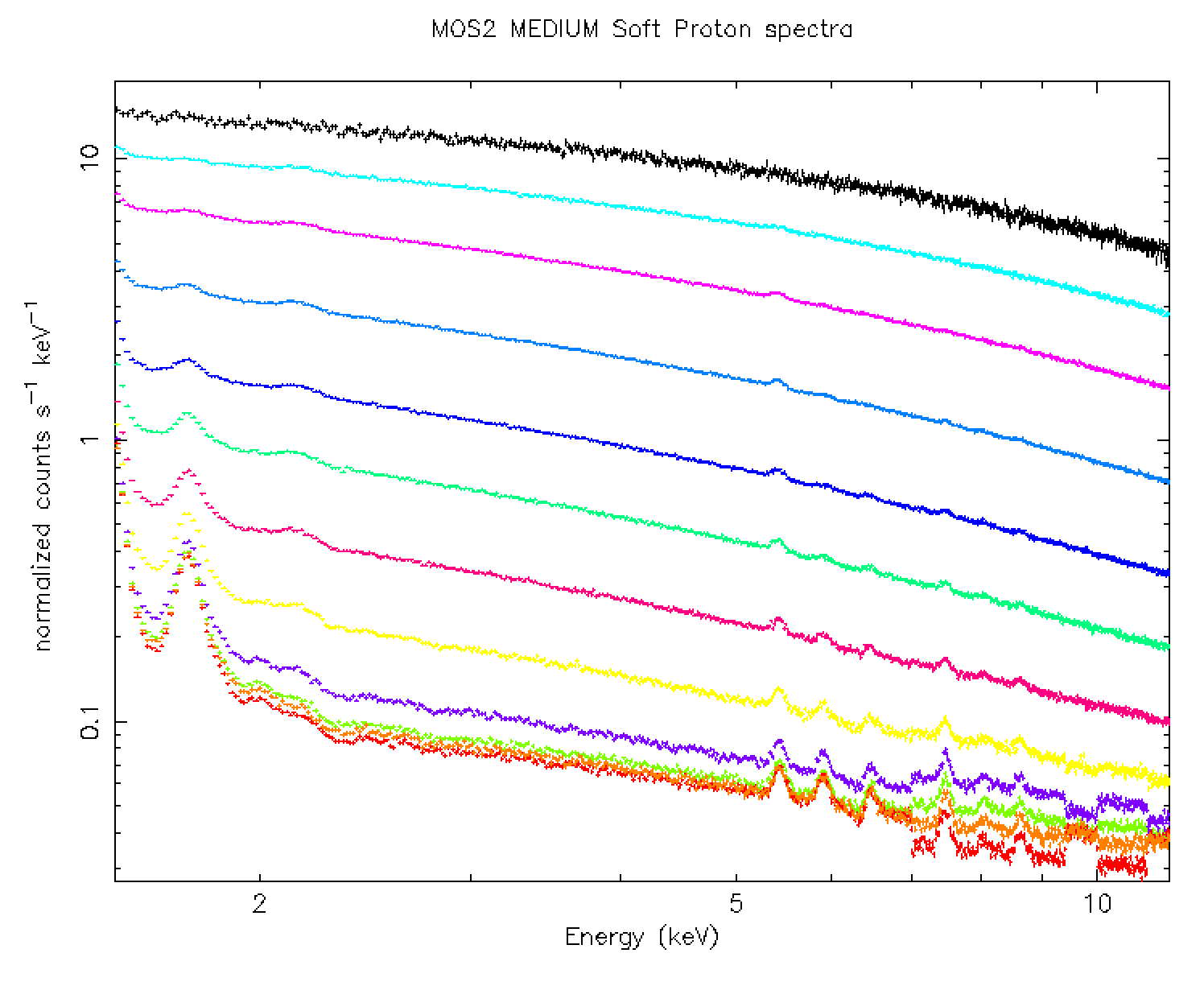}}
            {\includegraphics[width=0.87\hsize]{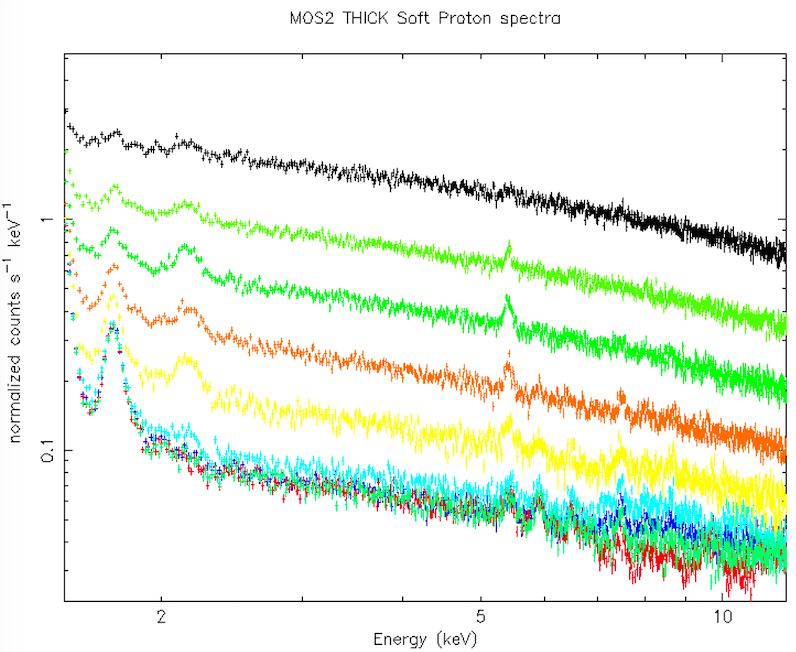}} 
      \caption{Set of MOS2 soft proton spectra extracted from different ranges of flare intensity of the EXTraS archive \citep{2017ExA....44..309S} for the thin (top), medium (middle), and thick (bottom) filters. The black spectra are the one selected for the validation.}
         \label{fig:13}
   \end{figure}
      \begin{figure}
   \centering
            {\includegraphics[width=0.88\hsize]{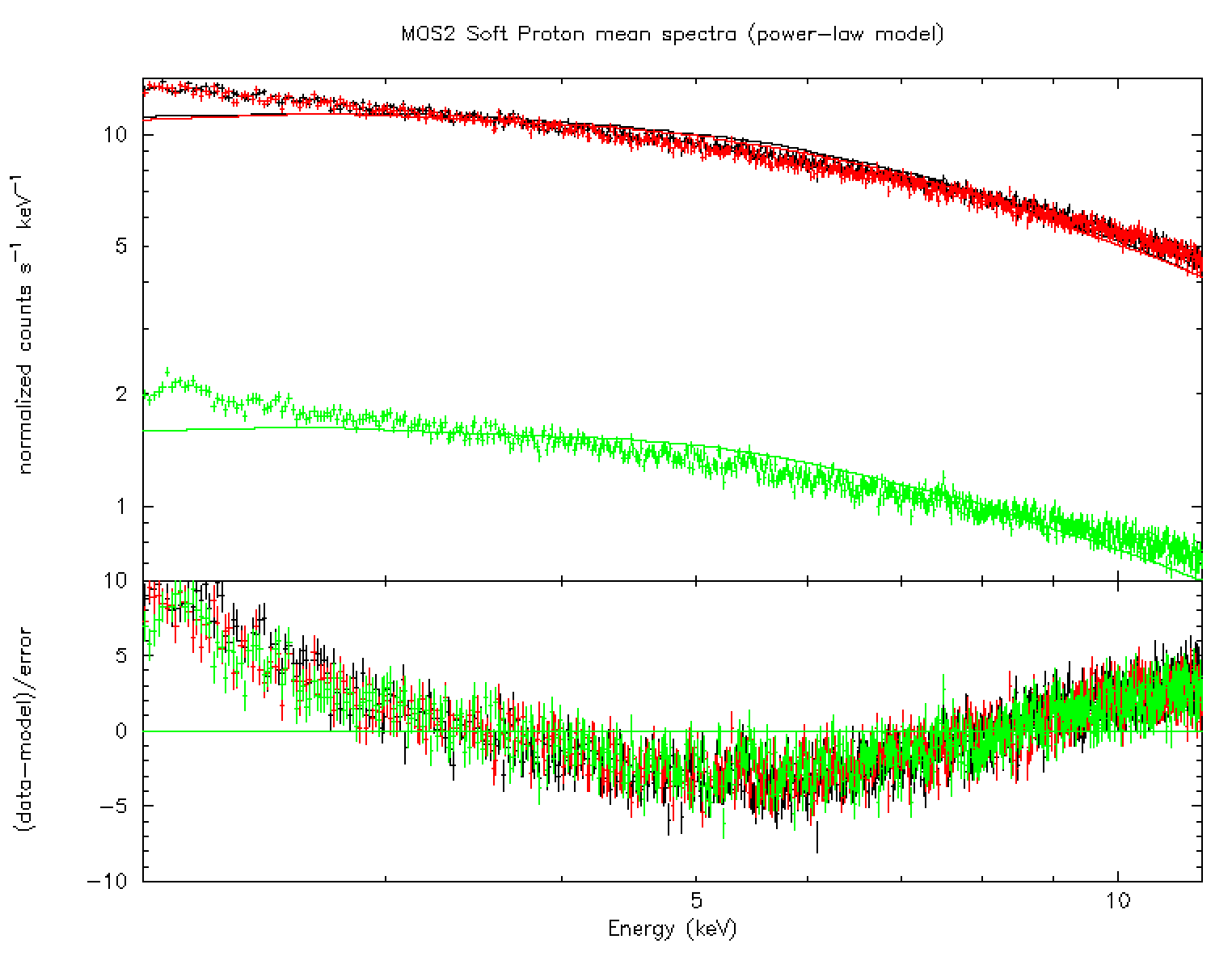}}
            {\includegraphics[width=0.9\hsize]{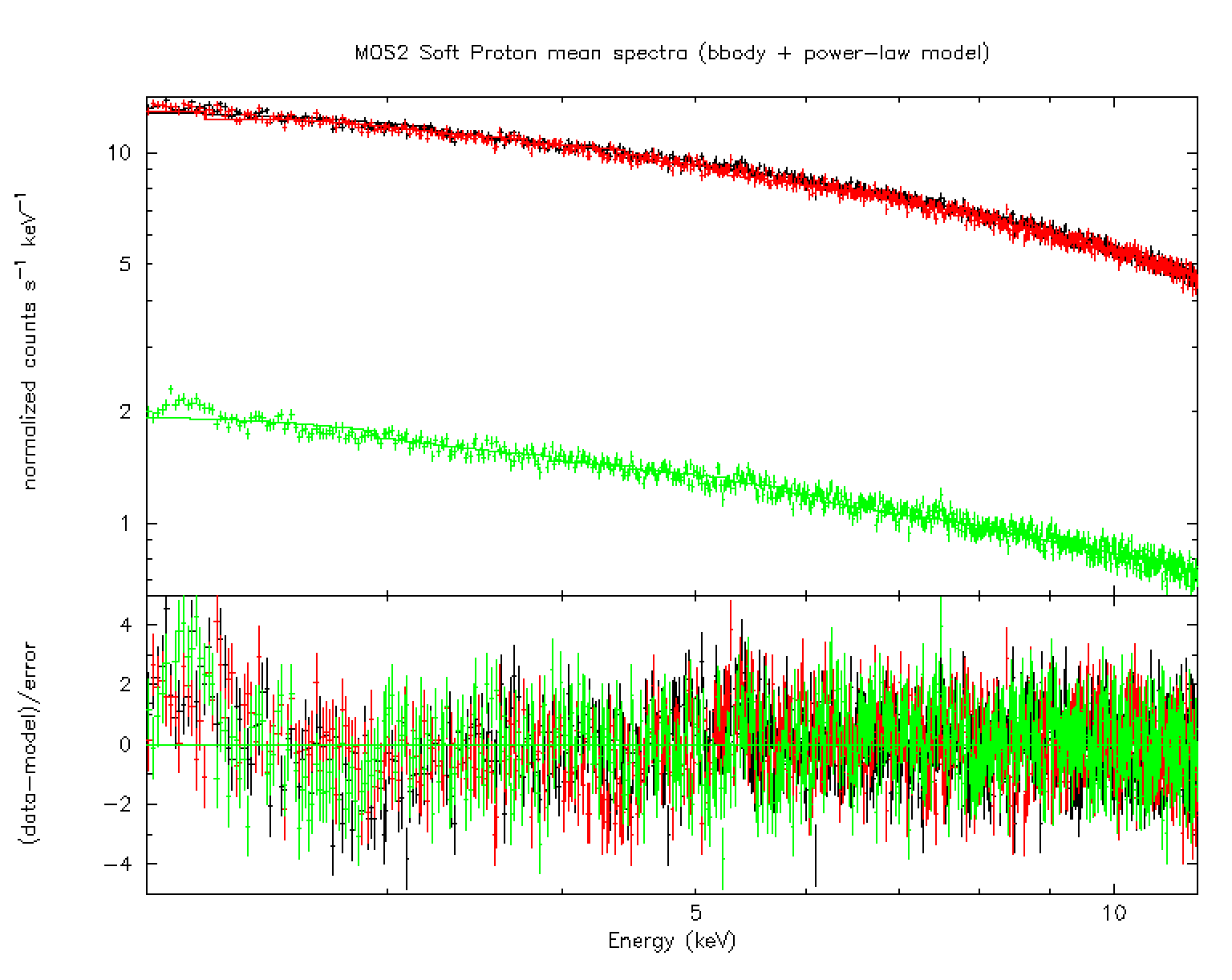}}
      \caption{Spectral analysis of the reference mean spectra using a power-law model (top) and a black-body + power-law model (bottom). The bottom panels show the residuals. The black, red, and green curves refer to the thin, medium, and thick filter configurations.}
         \label{fig:fit}
\end{figure}
The product of this analysis was a FITS file containing for each 500 sec time bin the count rate of both the in-FoV and out-of-FoV observations, for a total of 106.42 Ms of exposure \citep{AREMBES-MOLENDI}. The medium filter in-FoV minus out-of-FoV count rate cumulative distribution function (CDF) is normalised in units of counts cm$^{-2}$ s$^{-1}$ keV$^{-1}$, using the area coverage for the in-FoV observations of 31.2 cm$^{2}$ and the selected energy ranges in the data filtering process. 
\\
The CDF is shown in Fig. \ref{fig:cdf}, from which we can compute the maximum count rate at a given fraction of time spent by the \textit{XMM-Newton} telescope along its orbit. The maximum focused background count rate detected in 90\% of the observing time is 0.0164 counts cm$^{-2}$ s$^{-1}$ keV$^{-1}$, for the medium filter. The work of \citet{AREMBES-MOLENDI} also reported a mean count rate for the flaring background of 1.92, 1.48, and 0.74 cts/s, for the thin, medium, and thick filter. We used these values to scale the medium filter CDF 90\% count rate to the thin and thick filter configurations and perform the validation for the three MOS response files.
 \citet{2017ExA....44..309S} extracted 13 soft proton mean spectra from the EXTraS archive, for increasing ranges of flare intensity. They are shown\footnote{We excluded the 13th selection, with the highest intensity, because affected by large statistical fluctuations.} in Fig. \ref{fig:13} after subtracting the out-of-FoV from the in-FoV rates.  
For each filter configuration, we selected the highest spectrum (the black curve) to minimise the contamination by the instrumental background (for example the line peaks visible at different energies), and we normalised its intensity using the integral count rate obtained in the CDF 90\% to make it representative of the maximum flux expected in 90\% of exposure along the orbit.

\subsubsection{Spectral fitting}\label{sec:fit}

We fitted with \textit{Xspec} the reference soft proton spectra of Fig. \ref{fig:13}, selected above 2 keV, using the proton response matrix presented in this work. 
We first used a simple power-law model (PL), which is not able to explain the observed spectral distribution (Fig. \ref{fig:fit}, top panel). As studied in detail in Paper II, a simple power-law model is able to describe with sufficient accuracy the observed soft proton spectra above 5 keV, but leaves, on average, an excess below this energy of the order of 20\% for the MOS and 5\% for the pn. 
To investigate the origin of this excess, Paper II performed fits adding different components to the power law or using models with variable spectral index. They tested a double power-law model, a broken power-law, a power-law with an index which varies with energy as a log parabola (logpar), a power-law plus a black-body model.
Among these, acceptable $\chi^2$ values were only obtained adding a phenomenological black-body (BB) to a single power-law, which allows the full spectrum of the soft proton flare sample to be fit, while constraining the model free parameters.
The present analysis, applied to high statistics MOS2 spectra with averaged spectral distributions, confirms the results of Paper II. The addition of the black-body model is the only model that is able to describe with good accuracy the full spectrum, resulting in an overall $\chi^{2}$/d.o.f. of 2525/1896 and a null hypothesis probability of $1.5\times 10^{-21}$ (Fig. \ref{fig:fit}, bottom panel). The best-fit parameters, normalised for the intensity of the CDF at 90\%, are listed in Table \ref{tab:model} and refer to the model:
\begin{equation}
    M(E) = \frac{K_{BB}\times8.0525 E^{2}}{(kT^{4}[\rm exp(E/kT)-1])}\times K_{PL}E^{-\Gamma} \;.
\end{equation}
The residuals are in general below 2 $\sigma$ except below 3 keV where contamination induced features in the spectra are still present, in particular for the thick filter configuration where the proton flux is the lowest. We note that Paper II also finds positive residuals in some of the spectra below 3 keV with complex energy-dependent structures that a simple black-body can't explain.

\begin{table*}[!h]
\begin{center}
\caption{\label{tab:model}Best-fit parameters for a black-body + power-law model. }
\begin{tabular}{c|c|c|c|c|c}
\hline
\hline
\multirow{2}{*}{Filter} & \multirow{2}{*}{\rm K$_{\rm BB} (\times10^{7})$} & \multirow{2}{*}{\rm kT} & \multirow{2}{*}{\rm K$_{\rm PL} (\times10^{8})$} & \multirow{2}{*}{\rm $\Gamma$} & \multirow{2}{*}{$\chi^{2}/d.o.f$}\\
& & & & &\\
\hline
\multirow{2}{*}{Thin} & \multirow{2}{*}{$0.79_{0.53}^{1.18}$}  & \multirow{2}{*}{$1.31_{1.26}^{1.37}$} & \multirow{2}{*}{$0.50_{0.43}^{0.57}$} & \multirow{2}{*}{$3.30_{3.26}^{3.34}$} & \multirow{2}{*}{795/632}\\
& & & & & \\
\hline
\multirow{2}{*}{Medium}& \multirow{2}{*}{$2.55_{1.51}^{3.82}$}  & \multirow{2}{*}{$1.36_{1.31}^{1.43}$} & \multirow{2}{*}{$1.22_{1.01}^{1.38}$} & \multirow{2}{*}{$3.53_{3.48}^{3.56}$} & \multirow{2}{*}{783/632}\\
& & & & & \\
\hline
\multirow{2}{*}{Thick}& \multirow{2}{*}{$4.79_{2.41}^{10.63}$}  & \multirow{2}{*}{$1.88_{1.78}^{1.97}$} & \multirow{2}{*}{$0.77_{0.53}^{1.13}$} & \multirow{2}{*}{$3.37_{3.28}^{3.47}$} & \multirow{2}{*}{948/632}\\
& & & & & \\
\hline
\end{tabular}
\end{center}
\end{table*} 
\subsubsection{Results}\label{sec:val_res}
\begin{figure}
   \centering
   \resizebox{\hsize}{!}
            { \includegraphics{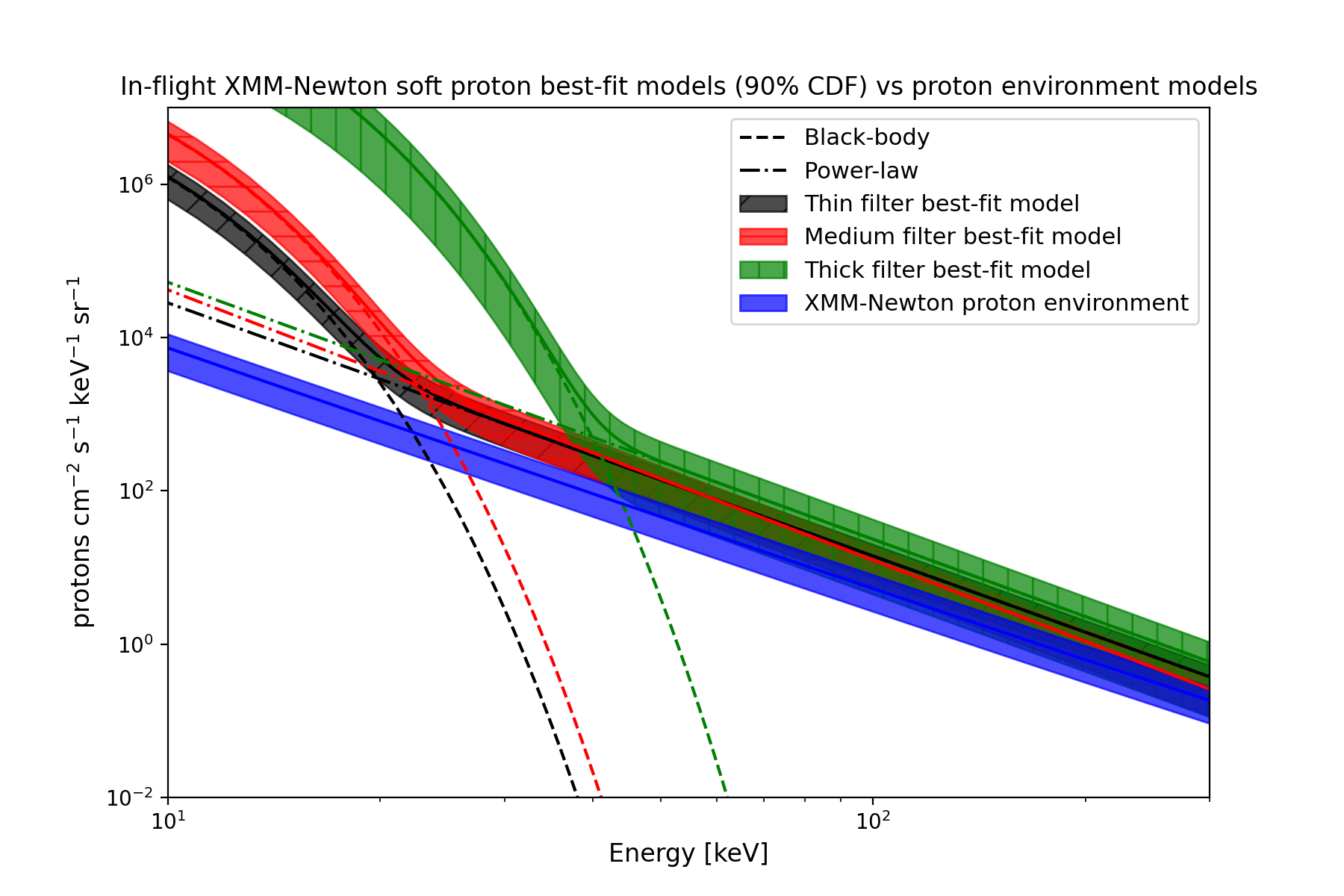}}
      \caption{Spectral models for the soft proton flares observed by \textit{XMM-Newton}, averaged for the maximum intensity observed in 90\% of the EXtraS archive, compared with the proton radiation environment predicted along its orbit for a cumulative fraction of 90\% (in blue). The black-body model used as a correction factor is shown with a dashed line.}
         \label{fig:val}
   \end{figure}

The best-fit spectral models for the observations obtained with the thin (in black), medium (in red), and thick (in green) filters are shown in Fig. \ref{fig:val}. The 68\% confidence error in the flux is summed to the systematic uncertainty of 30\% of the response files and applied to the models obtained from the present analysis. The model representative for the proton radiation environment encountered by \textit{XMM-Newton} (Sect. \ref{sec:input}) during the time covered by the EXtraS archive is plotted in blue. 
\\
The power-law component becomes dominant above a few tens of keV, and the slope for the three data-sets, relative to the different filters, is similar to the observed distribution of the protons trapped in the magnetosheath and nearby regions, where \textit{XMM-Newton} spent most of its time. This result is confirmed by the independent analysis of the three reference spectra. The soft proton flux extracted for a cumulative fraction of 90\% is also consistent, at the order of magnitude, with the mean proton intensity measured in the magnetosphere at 90\% CDF. 
In general, the power-law fluxes are higher than the predicted proton model. Since the latter is extracted in the magnetotail region at distances $>8$ Earth radii, farther than \textit{XMM-Newton} orbit, and XMM's soft proton rates are observed to reduce of a factor of few from the closest to the farthest distance from Earth \citep{2017ExA....44..273G}, we could explain such behaviour in terms of differences in \textit{Athena} and \textit{XMM-Newton} orbits. However, Paper II also finds a systematic increase of the analysed proton flux in the MOS CCDs to be a factor 2 higher than pn, likely caused by the MOS larger systematic excess below 5 keV. 
\\
The proton radiation environment measurements start from about 50 keV and are extrapolated at lower energies, so in principle a deviation from a power-law distribution, such as the modelled black-body-like distribution, could be justified by a secondary, very soft, background component. However, the analysis in Paper II shows that the low-energy excess modelled by the black-body correlates with the flux obtained from the power-law, that is, it origins from the same phenomenon. In addition, the black-body `contamination' increases with the amount of passive material between the mirror and the CCDs. We in fact find in Paper II a smaller contribution in the pn spectra, back-illuminated with no electron structure in front of the CCDs. In the present analysis the black-body component increases with the thickness of the optical filters, as clearly visible in Fig. \ref{fig:val}, and the difference among the three curves is similar to the simulated proton transmission of Fig. \ref{fig:rmf}. For all these reasons, we exclude the presence of an additional background component and no reliable physical information can be drawn from the black-body model. Instead we explain it as a correction factor required by the systematic overestimation of the simulated proton attenuation at low energies, that is counteracted by the black-body component. As previously discussed, the proton stopping power at low energies and for thin layers is still subject to large uncertainties. Also, the approximated mass model for the electrode structure in the front-illuminated MOS CCDs added a further layer of uncertainty in the proton energy loss. On-ground calibrations of the proton stopping power with EPIC-like passive materials are mandatory to improve the accuracy of the proton response files.

   \begin{figure*}
   \resizebox{\hsize}{!}
               {\includegraphics[width=0.3\linewidth]{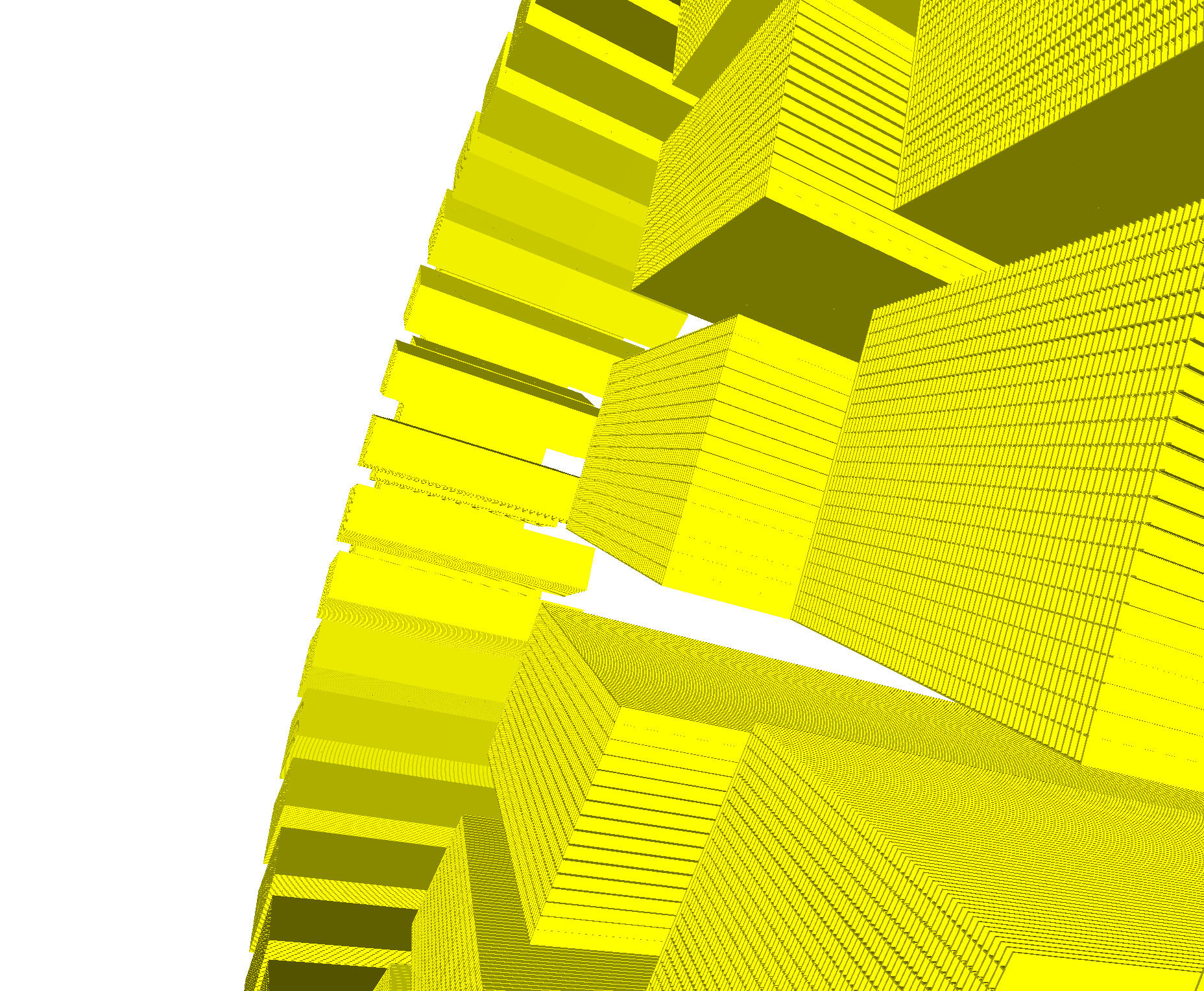}
            {\includegraphics[width=0.35\linewidth]{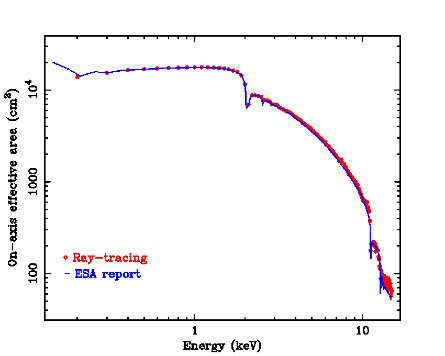}}
            {\includegraphics[width=0.35\linewidth]{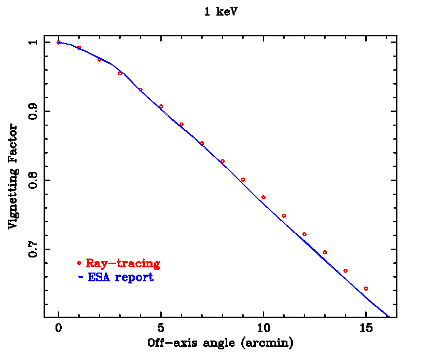}}
            }
      \caption{\textit{Athena} SPO simulation. (left) Close-up view of the mirror modules composing the Geant4 SPO mass model. SPO X-ray effective area (centre) and vignetting (right) simulated in ray-tracing (in red) and compared with the official results (in blue) reported in ATHENA - Telescope Reference Design and Effective Area Estimates, ESA-ATHENA-ESTEC-PL-DD001, Issue 3.3, 21/12/2020.}
         \label{fig:spo}
   \end{figure*}
           \begin{figure*}
      \centering
      \includegraphics[width=0.8\linewidth]{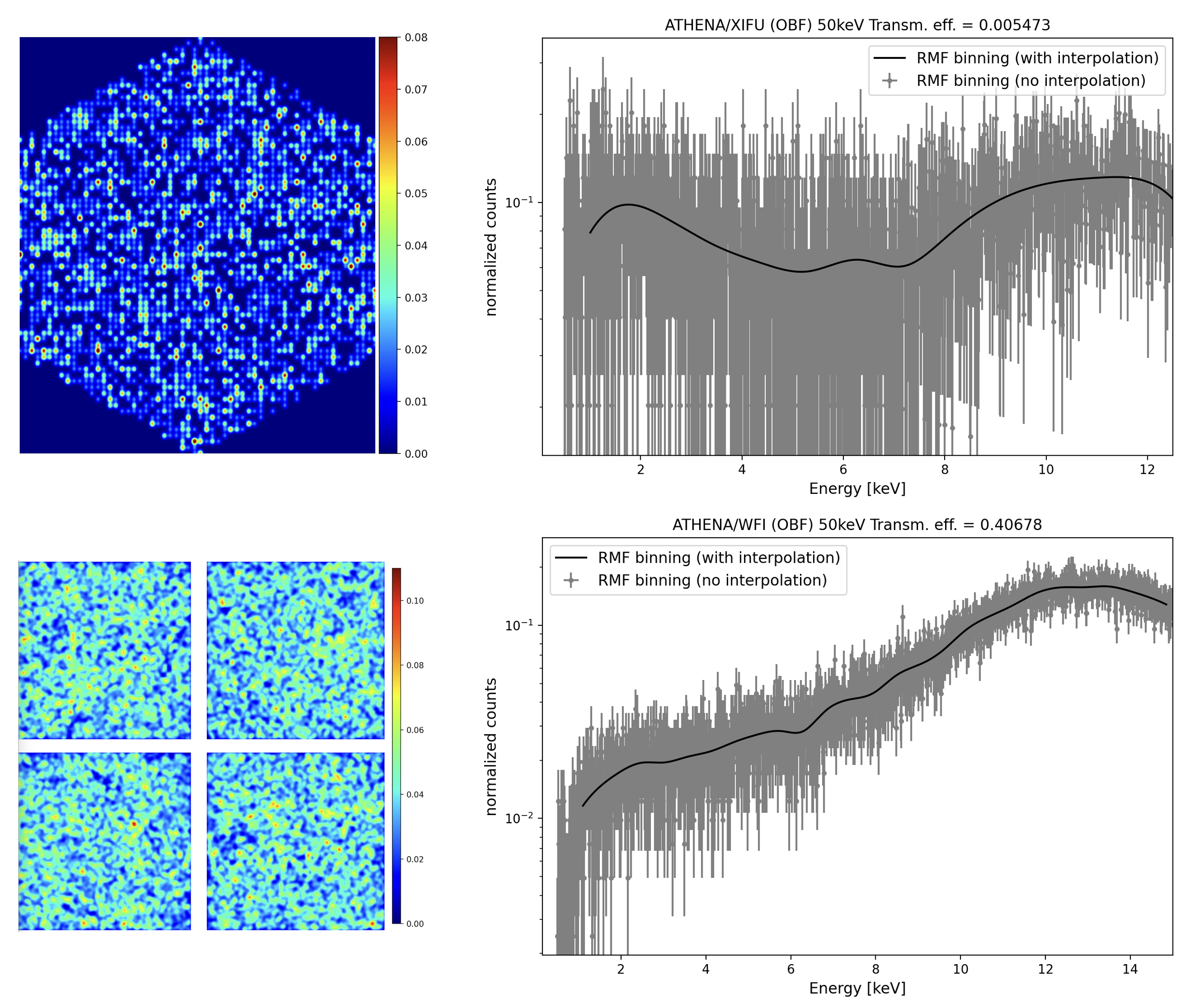}
      \caption{\textit{Athena} count map (left) and interpolation of the detected spectral distribution (right) for an input energy of 50 keV. The top and bottom panels refer to the X-IFU and WFI (with OBF) simulations.}
         \label{fig:athena_maps}
   \end{figure*}
\section{The \textit{Athena} proton response matrix}\label{sec:athena}

The \textit{Athena} space telescope was an X-ray observatory selected as the $2^{nd}$ ESA L-class X-ray mission \citep{2013arXiv1306.2307N} to fly in L1, designed to address the “The Hot and Energetic Universe” Cosmic Vision. In 2023 the nominal \textit{Athena} design was rescoped and its adoption by ESA moved to 2027, with a launch planned for the late 2030s. 
The core design stayed the same. The X-ray modular mirror is based on Silicon Pore Optics (SPO) technology \citep{2023SPIE12679E..02B} with a focal length of 12 m. Two instruments populate the focal plane, covering the soft < 15 keV energy range: a Wide Field Imager (WFI, \cite{2019SPIE11118E..0YM}) for wide field imaging and spectroscopy and an X-ray Integral Field Unit (X-IFU, \citet{2023ExA....55..373B}) for fine X-ray spectroscopy. 
\\
Geant4 simulations of the soft proton-induced background at both \textit{Athena} detectors \citep{Lotti2018, 2018ApJ...867....9F} showed that soft protons can threaten the achievement of the \textit{Athena} scientific requirements. For this reason, a magnetic diverter \citep{2018SPIE10699E..4AF} was designed to shield the focal plane from charged particles entering the field of view, and simulations for the focused background drove its specifications.
Because of the complexity and large diameter of the mirror, the Geant4 simulations of the proton scattering effect require significant CPU running times to get a minimum statistical level, so testing different input models and exploring different requirements for the diverter is challenging. Using the design of the old \textit{Athena} mirror and focal plane assembly, we built the proton response matrix to provide a quick evaluation of the soft proton-induced background level - without the magnetic diverter - to optimise the diverter design and better characterise the focused background. Since the new design foresees, among other changes, a reduced effective area from 1.4 to 1 m$^{2}$ at 1 keV, the results shown here represent an upper limit for the future mission.
 
\subsection{SPO proton scattering}\label{sec:spo}

The \textit{Athena} SPO proton scattering efficiency was extensively evaluated during the AREMBES project with independent mirror models built using Geant4 and ray-tracing simulation frameworks \citep{2018ApJ...867....9F}, with a systematic difference in the efficiency of about 20\% using the same scattering model. Later, a new SPO design was released \citep{2021SPIE11822E..04F} with an updated module layout and a rib pitch of 2.3 mm, still for a total of 15 rows of mirror modules. The new rib pitch was easily included in Geant4, but changing the layout of the mirror modules would have required to completely change the logic behind the Geant4 parameterised volumes that compose the mass model. We decided instead to only include the new layout in the raytracing simulation framework, and introduce its effect as a normalisation factor in the Geant4-based ARF. The raytracing simulation was verified by evaluating the on-axis X-ray effective area and vignetting and comparing it with the official performance results (Fig. \ref{fig:spo}, centre and right panels). In terms of efficiency, the new mirror layout increments the proton effective area by a 1.5 factor, while keeping the same spectral and angular distribution. This factor was added to the \textit{Athena} ARF files.

\subsection{X-IFU and WFI transmission efficiency}\label{sec:athena_fpa}
The Geant4 mass model of the X-IFU FPA was provided by the \textit{Athena} X-IFU instrument background working group and a detailed description can be found in \citet{2021ApJ...909..111L}. The instrument, operating in the 0.2 -- 12.5 keV energy range, is composed by thousands of transition-edge sensor (TES) microcalorimeters at the bottom of X-ray absorbers (Bi/Au) operated at 50 mK with a 249 $\mu$m pitch and 2.3 cm$^{2}$ of sensitive area. The mass model includes the set of fixed thermal filters placed within the cryostat thermal stages aperture at the top of the detector. We refer to the X-IFU filters as Optical Blocking Filters (OBF) for consistency with the WFI design. 
\\
The WFI FPA Geant4 mass model \citep{2018ApJ...867....9F} includes: a simplified wide field detector, composed of four pixelated quadrants each with $512\times512$, $130\times130$ $\mu$m$^{2}$ side pixels, with a thickness for the Si sensor of 450 $\mu$m; the fixed on-chip filter covering the pixels of the four quadrants and composed, from top to bottom, of 90 nm of Aluminium (Al), 30 nm of Silicon Nitride (Si3N4), and 20 nm of Silicon Oxide (SiO2); the squared $17\times17$ cm$^{2}$ Optical Blocking Filter on the filter wheel composed from top to bottom of 30 nm of Al and 150 nm of Polyimide. This filter is optional depending on the wheel position. The WFI energy range extends to 15 keV.
\\
The input protons used for the FPA are extracted for the X-IFU at the aperture cylinder entrance at 388 mm from the detector, for the WFI at the filter wheel at 10 cm from the detector, neglecting any secondary scattering with the Aluminium baffle on top of the filter wheel in favour of a larger statistics to model the energy redistribution. The counts map and X-ray spectrum obtained for an input energy of 50 keV is shown in Fig. \ref{fig:athena_maps}. As for the EPIC response files, the energy spectrum is first re-binned and then interpolated, and the interpolation function is used to fill the RMF.
 \begin{figure}
    \resizebox{\hsize}{!}
            {\includegraphics[width=\hsize]{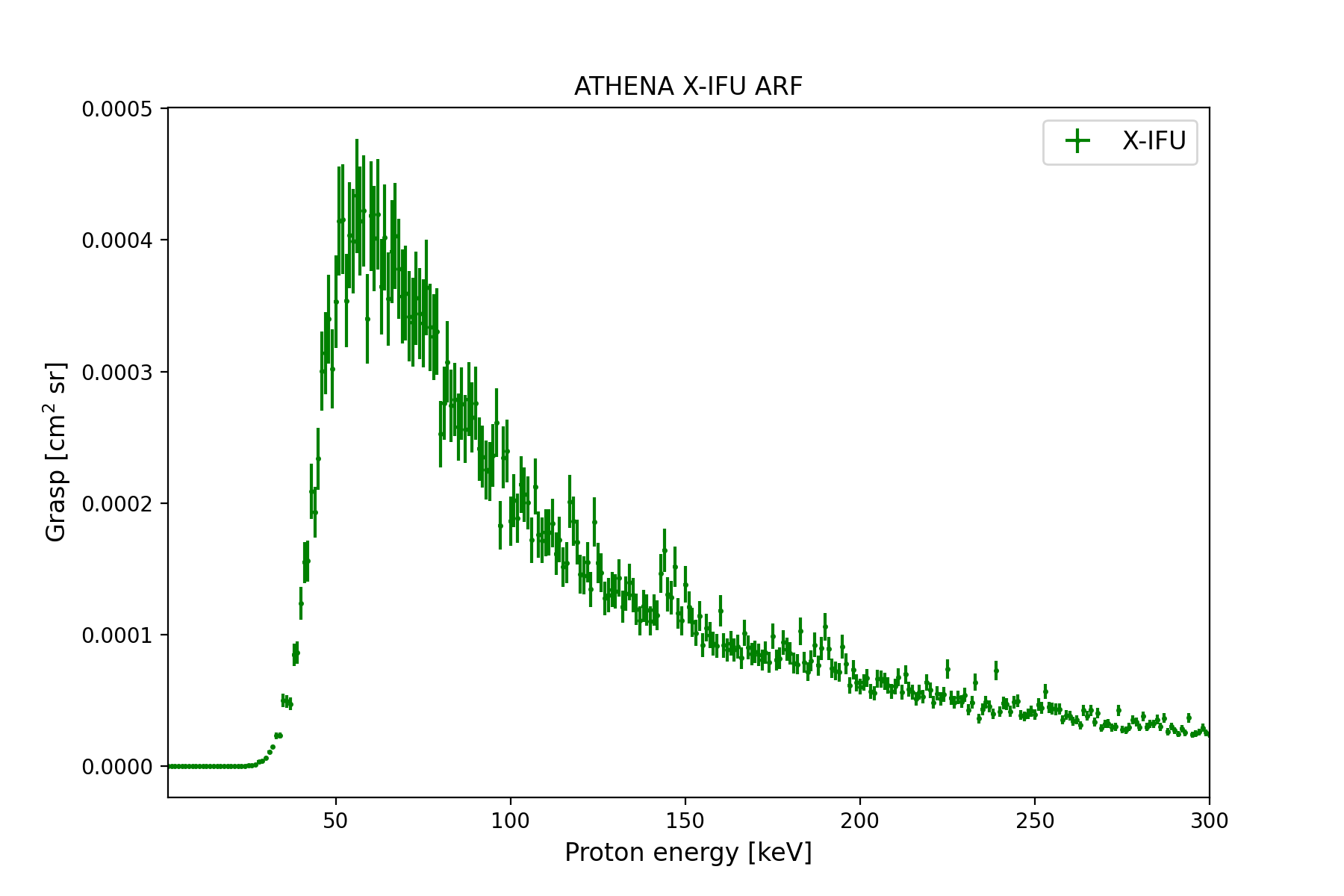}}     \\
            {\includegraphics[width=\hsize]{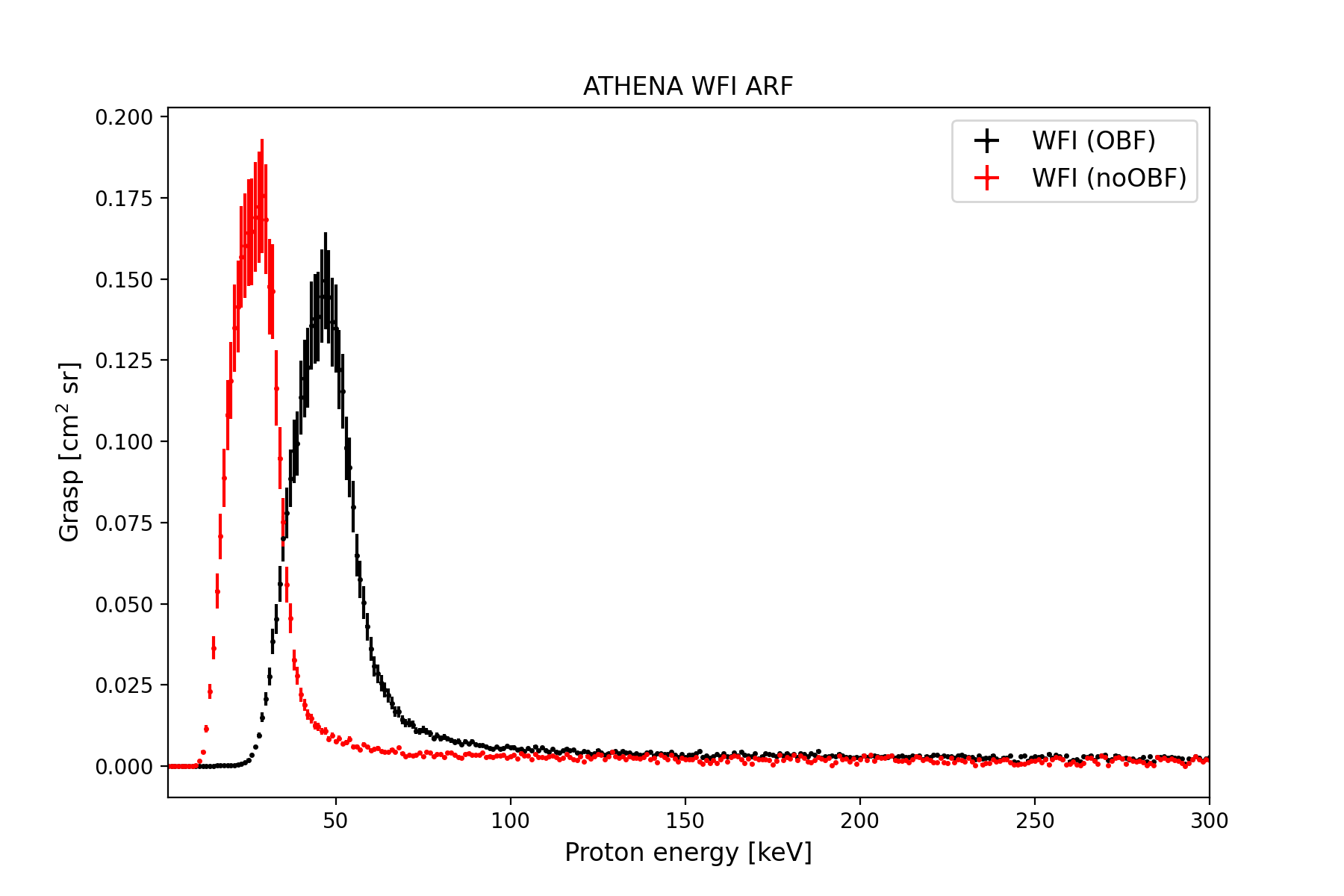}}
      \caption{Grasp curves stored in the ARF of the \textit{Athena} X-IFU (top) and WFI (bottom), including a 10\% systematic uncertainty.}
         \label{fig:athena_arf}
   \end{figure}
    \begin{figure}
    \resizebox{\hsize}{!}
            {\includegraphics[width=\hsize]{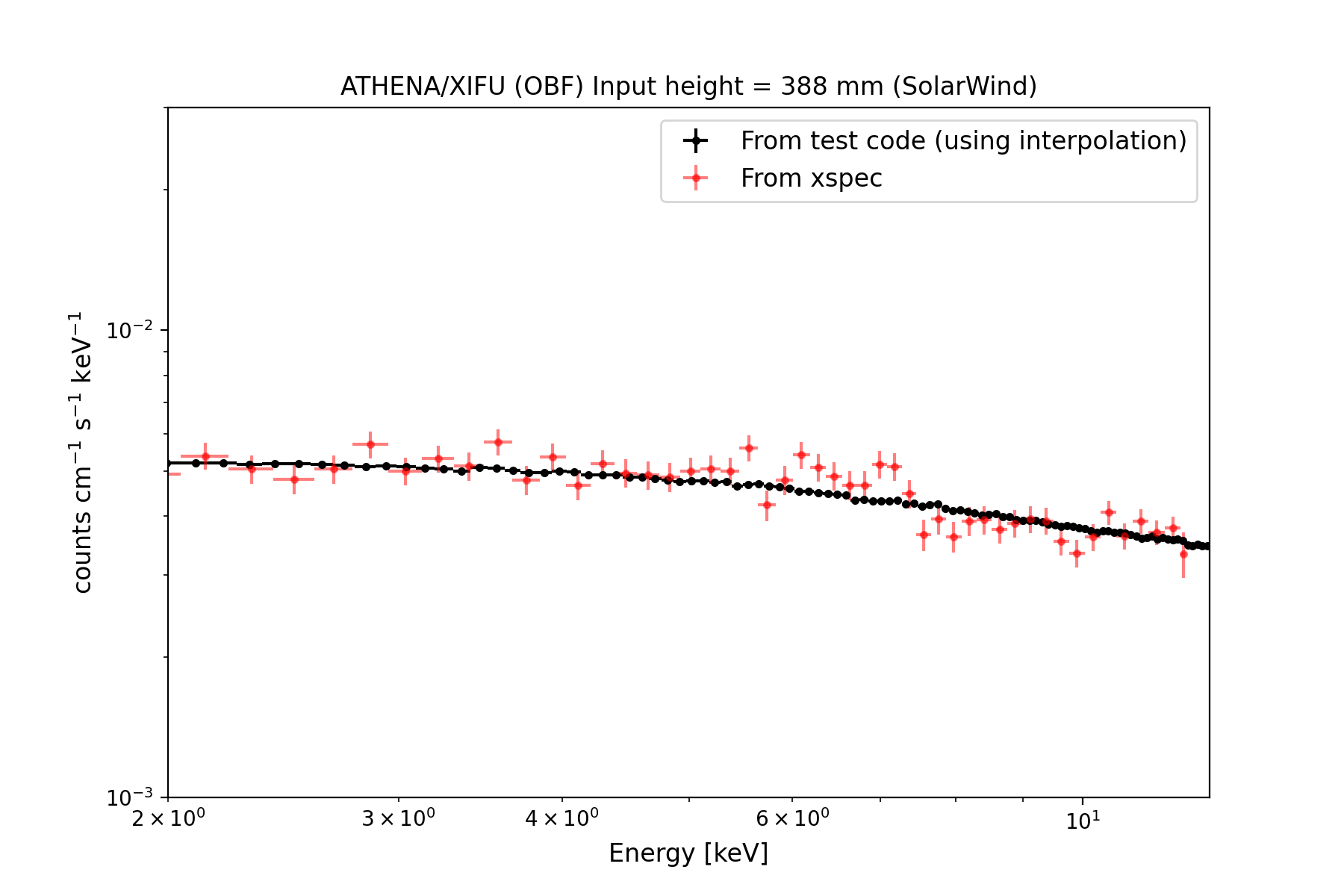}}     \\
            {\includegraphics[width=\hsize]{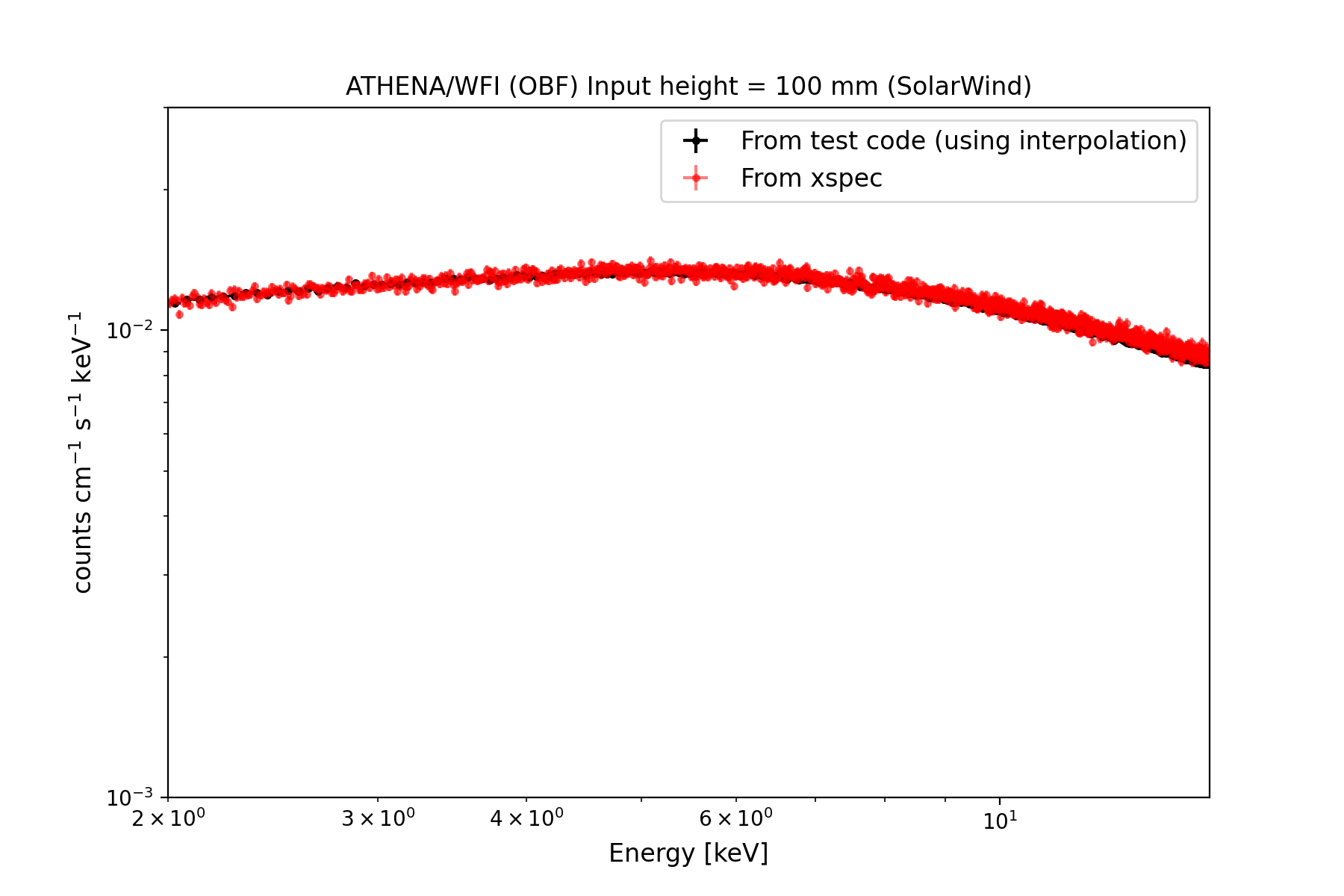}}
      \caption{Soft proton-induced background simulated with \textit{Xspec} (in red) and compared with a standard Geant4 simulation (in black), for the X-IFU (top) and WFI (OBF, bottom). The solar wind model at 90\% cumulative fraction was used in input.}
         \label{fig:athena_bkg}
   \end{figure}
\subsection{\textit{Athena} focused background}\label{sec:athena_resp}
The energy distribution of the WFI is binned according to the instrument X-ray channels, a total of 1485 channels, an energy width of 10 eV and energy boundaries of 0.15 - 15 keV. Because of the smaller aperture and detection area of the X-IFU, its simulation was the one that suffered the most from the limited statistics and using the original energy resolution ($<2.5$ eV up to 7 keV), the smallest among the simulated detectors, was unfeasible for the available computing resources. A binning factor of 25 was used instead, for a total of 1196 channels, an increasing energy width from about 5 eV to about 30 eV and energy boundaries of 0.06116 - 12.49961 keV. 
The simulated grasp stored in the ARF files is shown in Fig. \ref{fig:athena_arf}. The X-IFU proton detection efficiency extends with a long tail up to 300 keV. This is likely caused by multiple secondary scattering within the aperture cylinder, with protons losing energy at each scattering with the inner surface. The proton transmission for the WFI peaks at about 30 keV if no OBF is placed in the line of sight, and at about 50 keV with the OBF.
\begin{figure}
    \resizebox{\hsize}{!}
            {\includegraphics[width=\hsize]{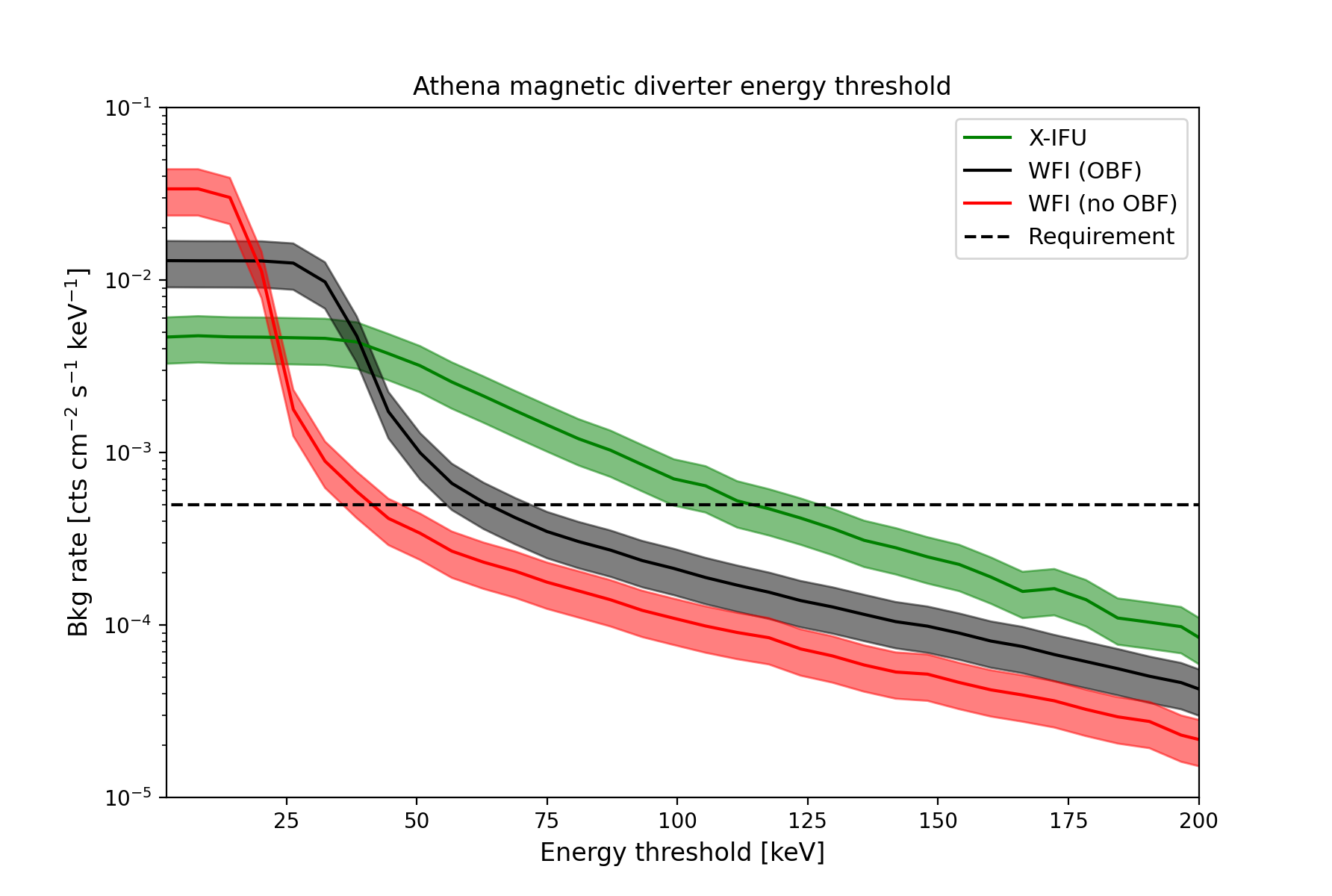}} 
      \caption{Mean focused backgrond rate of X-IFU and WFI as a function of the energy threshold applied by a magnetic diverter.}
         \label{fig:md}
\end{figure}
\\
As for \textit{XMM-Newton}, we selected only single pixels events for building the response files. The validity energy ranges for their use, required to avoid artefacts produced by the interpolation model, are 2 - 11.5 keV for the WFI. For the X-IFU, from 2 to 7 keV. The limited upper energy threshold is due to the lower statistics achieved in the X-IFU Geant4 simulations. We predict, as for \textit{XMM-Newton}, an overall systematic uncertainty of 30\%, dominated by the SPO mass model approximations, about 20\%, and the Geant4 intrinsic limitations in modelling low-energy proton interactions within thin layers, about 10\%. 
\\
The \textit{Athena} response matrix was verified by comparing the \textit{Xspec} simulation, using the RMF and ARF files, with a standard Geant4 simulation, using the L1 solar wind proton spectrum (90\% CDF) in input, shown in Fig. \ref{fig:athena_bkg}. The poor statistics of the X-IFU simulations limited the accuracy of the energy interpolation: while the spectrum is in general agreement with the reference simulation, artefacts are visible in the X-ray spectrum. A dedicated simulation campaign, with extended computational resources, would solve the issue and would also extend the validity range of the response files.
\\
The simulated spectra are also updated prediction of the soft proton-induced background for the old \textit{Athena} mission, using both an updated SPO mirror design and the X-IFU FPA mass model with respect to the AREMBES studies and based on a validated simulation pipeline, the same as the one used for the \textit{XMM-Newton} response files. Given a focused background flux requirement of $5\times10^{-4}$ cts cm$^{-2}$ s$^{-1}$ keV$^{-1}$ in the 2 -- 7 keV (WFI) and 2 -- 10 keV (X-IFU) energy ranges in 90\% of operational time, present results confirm that soft proton flares would have contaminated the observations of the nominal \textit{Athena} mission. A magnetic diverter would deflect the protons outside the focal plane, with an angle that only depends on the proton energy. Depending on the magnetic intensity, the effect of the diverter is a low-energy cut to the protons reaching the detectors. Using \textit{Xspec} and the \textit{Athena} response files, we computed the residual focused background as a function of the diverter energy threshold, for a solar wind input model. The result is shown in Fig. \ref{fig:md}, with a 30\% systematic uncertainty applied to the simulated background. The energy threshold to ensure a background below the requirement is about 70 keV for the WFI and 120 keV for the X-IFU, the latter being higher because of the higher peak and extended tail of the effective area. As noted earlier, such results must be considered as upper limits in view of the updated \textit{Athena} design.

\section{Conclusions}\label{sec:sum}
We built a verified end-to-end Geant4 simulation of the \textit{XMM-Newton} proton flares, from the scattering with the X-ray mirror to the transmission through the optical filters, the MOS on-chip electrode and the final absorption in the EPIC CCDs. The resulting energy redistribution and proton detection efficiency are encoded into NASA caldb RMF and ARF response files in FITS format.  Any X-ray data analysis tool available to the X-ray astronomy community and compliant with the NASA caldb format can be used to simulate the soft proton-induced background spectra, for any given condition of the orbit proton environment without the need to run again the simulation pipeline. The same tools can be used to fit the in-flight soft proton spectra and model the flux and energy distribution of the proton environment along the orbit. We chose the NASA \textit{Xspec} fitting package for the verification and validation activity. The response files are verified by comparing the simulated background, using  \textit{Xspec}, with independent results obtained with a standard simulation pipeline. Mean in-flight soft proton spectra, representative of the maximum flux expected in 90\% of operational time, are fitted with the response files, and the best-fit model is compared with independent measurements of the proton radiation environment in the magnetosheath and nearby regions. Above a few tens of keV, the best-fit power-law distribution is in good agreement with the predicted proton flux along \textit{XMM-Newton} orbit, proving the validity of the simulation chain. At lower energy, a black-body-like emission is required as a correction factor for the systematic error in the proton transmission simulation. The cross-correlation of simultaneous MOS and PN observations carried out in Paper II leads to an overall systematic accuracy of the extracted input proton fluxes within a factor 2.
\\
The same simulation pipeline, but using the \textit{Athena} Silicon Pore Optics and the X-IFU and WFI X-ray instruments, is used to produce a proton response file for the old, before rescoping, mission proposal and update the prediction for the focused charged background.
\\
The ability to disentangle the soft proton contamination from the X-ray astrophysical source would allow tens and more Megaseconds of observations to be recovered and would represent a major breakthrough in the \textit{XMM-Newton} data analysis. While this remains a use case for the proton response matrix, the systematic uncertainties in the simulation chain together with the aleatory nature of the soft proton spectral distribution would cause a large degeneracy with the models for the X-ray target. 
\\
Dedicated on-ground calibration campaigns, that we are currently lacking, for the proton response matrix would increase its accuracy and its range of applications. It's been proved that the unfocused NXB of \textit{Chandra} and \textit{XMM-Newton} \citep{2017ExA....44..321G, 2022SPIE12181E..2EG} correlates with simultaneous measurements of the cosmic-ray flux, and this behaviour can be exploited to predict the same type of background in future X-ray missions. 
The work of \citet{2017ExA....44..321G} also showed that such correlation is not present for soft protons when comparing their flares with low-energy proton monitors (for instance the ACE Satellite in L1), likely due to subsequent interactions in the Earth's magnetosphere that alter their path and intensity. However, recent studies using a machine learning approach trained on Cluster/RAPID observations \citep{2021ApJ...921...76K} showed that it is possible to predict the soft proton intensity at different points of the Earth's magnetosphere and potentially build a proton model tailored for the \textit{XMM-Newton} orbit. The availability of a local model for the proton environment would not only extend the validation process of the response matrix, but also predict hence potentially remove the focused background.
\\
Despite such limitations and thanks to a thorough verification process, we were able for the first time to analyse soft proton flares with \textit{Xspec}, converting an X-ray grazing telescope into a proton monitor. Comparing the best-fit model with independent measurements of the radiation environment, we definitively proved the origin of the background flares to be magnetically trapped low-energy protons. In addition, we proved the validity of the proton scattering simulations to drive the design of dedicated shielding solutions, for example a magnetic diverter, for future missions.
Additional studies will explore in more detail the proton flux for different orbital regions, hopefully together with more refined measurements from space radiation monitors. 
\\
The first release for the proton RMF and ARF files of \textit{XMM-Newton} and \textit{Athena} is publicly distributed in
\href{https://zenodo.org/records/7629674}{Zenodo}, while the latest releases and updates can be found in the \href{https://www.ict.inaf.it/gitlab/proton_response_matrix}{INAF Gitlab project}.

\begin{acknowledgements}
     The research leading to these results has received funding from the European Union’s Horizon 2020 Programme under the AHEAD2020 project (grant agreement n. 871158), the Accordo Attuativo ASI-INAF n. 2019-27-HH.0 and the EXACRAD - CTP ESA contract n. 4000121062/17/NL/LF.
\end{acknowledgements}

%
%

\bibliographystyle{aa}       
\bibliography{xmm_sp_response_1}   
\end{document}